\documentclass{aa}  
\usepackage{graphicx}
\usepackage[varg]{txfonts}
\usepackage{subfig}
\usepackage{newtxtext,newtxmath}
\usepackage[labelfont=bf]{caption}
\usepackage{longtable,booktabs}
\usepackage{pdfpages}

\DeclareMathOperator{\sech}{sech}

\newcommand\Eq{Eq.\ }
\newcommand\Eqs{Eqs.\ }
\newcommand\de{\partial}

\newcommand\alphaone{\alpha_1}

\newcommand\betaone{\beta_1}
\newcommand\bone{b_1}

\newcommand\Eone{E_1}

\newcommand\Azero{A_0}
\newcommand\Etwo{E_2}
\newcommand\alphatwo{\alpha_2}
\newcommand\betatwo{\beta_2}
\newcommand\Fgas{F_{\rm gas}}
\newcommand\Fstar{F_\star}
\newcommand\Fone{F_1}
\newcommand\Ftwo{F_2}
\newcommand\Foneplustwo{F_{1+2}}
\newcommand\alphagas{\alpha_{\rm gas}}
\newcommand\alphastar{\alpha_\star}

\newcommand\Fgasplusstar{F_{\rm gas+\star}}

\newcommand\Sigmai{\Sigma_i}

\newcommand\Sigmagas{\Sigma_{\rm gas}}
\newcommand\Sigmastar{\Sigma_\star}
\newcommand\sigmastarR{\sigma_{\star,R}}
\newcommand\sigmastarz{\sigma_{\star,z}}
\newcommand\Sigmagastilde{\tilde\Sigma_{\rm gas}}
\newcommand\Sigmastartilde{\tilde\Sigma_\star}
\newcommand\sigmagas{\sigma_{\rm gas}}
\newcommand\sigmaturb{\sigma_{\rm turb}}
\newcommand\rhozerogas{\rho_{\rm 0,gas}}
\newcommand\rhozerostar{\rho_{0,\star}}
\newcommand\rhogas{\rho_{\rm gas}}
\newcommand\pressgas{\press_{\rm gas}}
\newcommand\pressstar{\press_\star}
\newcommand\rhostar{\rho_\star}
\newcommand\bgas{b_{\rm gas}}

\newcommand\zhalf{z_{\rm half}}
\newcommand\zinst{z_{\rm inst}}
\newcommand\zhalfgas{z_{\rm half,gas}}
\newcommand\zhalfstar{z_{\rm half,\star}}
\newcommand\zinstgas{z_{\rm inst,gas}}
\newcommand\zinststar{z_{\rm inst,\star}}
\newcommand\sigmastar{\sigma_\star}

\newcommand\kR{k_R}
\newcommand\kRtilde{\tilde{k}_R}
\newcommand\kB{k_{\rm B}}

\newcommand{\hz}{h_z}
\newcommand{\hzi}{h_{z,i}}
\newcommand{\hzone}{h_{z,1}}
\newcommand{\hztwo}{h_{z,2}}
\newcommand{\hzgas}{h_{z,{\rm gas}}}
\newcommand{\hzstar}{h_{z,\star}}
\newcommand{\hzgastilde}{\tilde{h}_{z,{\rm gas}}}
\newcommand{\hzstartilde}{\tilde{h}_{z,\star}}
\newcommand\deltarho{\delta \rho}
\newcommand{\beq}{\begin{equation}}
\newcommand{\eeq}{\end{equation}}
\newcommand{\imaginary}{{\rm i}}
\newcommand{\im}{\imaginary}
\newcommand\Phiext{\Phi_{\rm ext}}
\newcommand\Phihat{\hat\Phi}
\newcommand\Phitilde{\tilde\Phi}
\newcommand\Phihatzero{\hat\Phi_0}

\newcommand\Phione{\Phi_1}
\newcommand\Phitwo{\Phi_2}
\newcommand\Phii{\Phi_i}
\newcommand{\qunp}{q_{\rm unp}}
\newcommand\rhohat{\hat\rho}
\newcommand\rhohatzero{\hat\rho_0}
\newcommand\dd{{\rm d}}
\newcommand{\hseventy}{h_{70\%}}
\newcommand\ztilde{\tilde{z}}

\newcommand\rhogastilde{\tilde{\rho}_{\rm gas}}
\newcommand\rhostartilde{\tilde{\rho}_\star}
\newcommand\rhozero{\rho_0}
\newcommand\rhoone{\rho_1}
\newcommand\rhotwo{\rho_2}
\newcommand\rhoi{\rho_i}

\newcommand\rhozeroone{\rho_{0,1}}
\newcommand\rhozerotwo{\rho_{0,2}}
\newcommand\rhozeroi{\rho_{0,i}}
\newcommand\cs{c_{\rm s}}
\newcommand\csiso{\sigma}

\newcommand\sigmai{\sigma_i}

\newcommand\sigmaone{\sigma_1}
\newcommand\sigmatwo{\sigma_2}
\newcommand\csone{\sigma_1}
\newcommand\cstwo{\sigma_2}

\newcommand{\Qthreed}{Q_{\rm 3D}}
\newcommand{\Qthreedone}{Q_{\rm 3D,1}}
\newcommand{\Qthreedtwo}{Q_{\rm 3D,2}}
\newcommand{\Qthreedgas}{Q_{\rm 3D,gas}}
\newcommand{\Qthreedstar}{Q_{\rm 3D,\star}}

\newcommand{\Qgas}{Q_{\rm gas}}
\newcommand{\Qi}{Q_i}
\newcommand{\Qstar}{Q_\star}

\newcommand\mbar{\bar{m}}

\newcommand\press{p}
\newcommand\pressi{p_i}
\newcommand\presszero{p_0}
\newcommand\pressz{\press^{\prime}_z}
\newcommand\rhoz{\rho^{\prime}_z}
\newcommand\presszi{\press^{\prime}_{z,i}}
\newcommand\rhozi{\rho^{\prime}_{z,i}}
\newcommand\rhozone{\rho^{\prime}_{z,1}}
\newcommand\rhoztwo{\rho^{\prime}_{z,2}}

\newcommand\gammap{{\gamma^{\prime}}}
\newcommand{\uvi}{{\boldsymbol u}_i}
\newcommand{\uphii}{u_{\phi,i}}
\newcommand{\uphione}{u_{\phi,1}}
\newcommand{\uphitwo}{u_{\phi,2}}
\newcommand{\uRi}{u_{R,i}}
\newcommand{\uzi}{u_{z,i}}

\usepackage[]{hyperref}
\hypersetup{colorlinks=true,linkcolor=red,citecolor=blue,urlcolor=magenta}

\defcitealias{Nip23}{N23}

\begin{document}

\title{Local gravitational instability of two-component thick discs in
  three dimensions}

\authorrunning{C. Nipoti et al.}
\author{Carlo Nipoti~\inst{1}, Cristina Caprioglio\inst{1} and  Cecilia Bacchini\inst{2} } 

\institute{
1 Dipartimento di Fisica e Astronomia ``Augusto Righi'', Universit\`{a} di Bologna, via Gobetti 93/2, 40129, Bologna, Italy\\
\email{carlo.nipoti@unibo.it}\\
2 DARK, Niels Bohr Institute, University of Copenhagen, Jagtvej 155, 2200 Copenhagen, Denmark \label{inst:dark}\\
}

\date{Accepted, May 20, 2024}

 
\abstract{}{ The local gravitational instability of rotating discs is
  believed to be an important mechanism in different astrophysical
  processes, including the formation of gas and stellar clumps in
  galaxies. We aim to study in three dimensions the local
  gravitational instability of two-component thick discs.}  {We take
  as starting point a recently proposed analytic three-dimensional
  (3D) instability criterion for discs with non-negligible thickness
  which takes the form $\Qthreed<1$, where $\Qthreed$ is a 3D version
  of the classical 2D Toomre $Q$ parameter for razor-thin discs. Here
  we extend the 3D stability analysis to two-component discs,
  considering first the influence on $\Qthreed$ of a second
  unresponsive component, and then the case in which both components
  are responsive. We present the application to two-component discs
  with isothermal vertical distributions, which can represent, for
  instance, galactic discs with both stellar and gaseous
  components. {Finally, we relax the assumption of vertical
    isothermal distribution, by studying one-component
    self-gravitating discs with polytropic vertical distributions for
    a range of values of the polytropic index corresponding to
    convectively stable configurations}. }{We find that $\Qthreed<1$,
  where $\Qthreed$ can be computed from observationally inferred
  quantities, is a robust indicator of local gravitational
  instability, depending only weakly on the presence of a second
  component and on the vertical gradient of temperature or velocity
  dispersion. We derive a sufficient condition for local gravitational
  instability in the midplane of two-component discs, which can be
  employed when both components have $\Qthreed>1$.}{}

\keywords{
  galaxies: kinematics and dynamics -- galaxies: star
  formation -- instabilities -- planets and satellites: formation --
  protoplanetary discs -- stars: formation}

   \maketitle

 \section{Introduction}
 \label{sec:intro}
 
The local gravitational instability of rotating fluids is an important
phenomenon in several astrophysical systems, ranging from
protoplanetary discs to galactic discs.  In the case of razor-thin
rotating discs, \citet{Too64} derived an analytic instability
criterion against small axisymmetric perturbations.  However,
observations of astrophysical systems showed that the assumption of
infinitesimally thin discs is often not realistic. For instance, both
gaseous and stellar discs in nearby star-forming galaxies have
non-negligible thickness of $\sim 0.1-1$ kpc
\citep[e.g.][]{OBr10,van11,Yim14,Mac17,Mar17}.  Different
modifications to 2D Toomre's analysis that approximately account for
the finite thickness of the disc have been proposed in several works
\citep[e.g.][]{Too64,Van70,Rom92,Ber10,Wan10,Elm11,Gri12,Rom13,Beh15}.

In many cases, astrophysical discs are not purely self-gravitating,
but belong to multi-component systems.  A typical case is that of
late-type galaxies, in which gaseous discs coexist with stellar discs and with
pressure supported components like bulges and dark-matter halos.  When
two or more discs coexist, the local gravitational stability of each
disc can be influenced by the interplay with the other disc, because
any gravitational perturbation affect both discs at the same time
\citep{Too64}.  The stability of multi-component discs has been widely
studied in the literature in the case of infinitesimally thin discs
and also in 2D analyses that account approximately for the finite disc
thickness
\citep{Lin66,Kat72,Jog84bis,Jog84,Rom94,Wan94,Elm95,Raf01,She03,Rom11}.

\citet[][hereafter \citetalias{Nip23}]{Nip23} presented a
three-dimensional (3D) stability analysis of rotating stratified
fluids, which generalizes previous 3D studies that obtained
instability criteria based on more restrictive assumptions
\citep{Saf60,Cha61,Gol65a,Gol65b,Gen75,Ber82,Mam10,Mei22}.  Here we
extend the work of \citetalias{Nip23}, by studying in 3D the local
gravitational instability of two-component discs.  We first consider
the influence on the 3D instability parameter of a second unresponsive
component, and then the case in which both components are
responsive. We present the application to two-component discs with
isothermal vertical distributions, which can represent, for instance,
galactic discs with both stellar and gaseous components. Finally, we
relax the assumption of vertical isothermal distribution, by studying
self-gravitating discs with vertical polytropic distribution, and thus
in general with non-zero vertical temperature or velocity-dispersion
gradients.  {Though the computation of two-component discs with
  generic polytropic vertical distributions is not more technically
  difficult than that of of two-component isothermal discs, in this
  work we present results on the polytropic case only for
  one-component discs, which is a clean case study that allows us to
  quantify the effect of non-isothermality, without exploring the
  larger parameter space of two-component systems.}

The paper is organized as follows. Section~\ref{sec:prel} recalls some
relevant equations of previous disc stability analyses. In
section~\ref{sec:twocompisoth} we describe the properties of
two-component disc models with vertical isothermal distributions.  The
question of the linear stability in 3D of two-component discs with
both components responsive is addressed Section~\ref{sec:stabresp}.
Disc with vertical polytropic distributions are discussed in Section
\ref{sec:poly} and Section~\ref{sec:concl} concludes.

\section{Preliminaries} 
\label{sec:prel}

In this work we study axisymmetric rotating fluid disc models,
adopting, as it is natural, a cylindrical system of coordinates
$(R,z,\phi)$. As we limit our stability analysis to axisymmetric
perturbations, none of the quantities considered in this paper depends
on $\phi$.  Let us consider the properties of a disc at given radius
$R$. The disc surface density is
\begin{equation}
\Sigma=\int_{-\infty}^\infty\rho(z)\dd z,
\label{eq:surf_den}
\end{equation}
where $\rho(z)$ is vertical volume density profile.  We assume that
$\rho(z)$ is such that $\Sigma$ is finite, and we introduce as a
reference vertical scale the half-mass half-height $\zhalf$, defined to be
such that
\begin{equation}
\frac{1}{\Sigma}\int_{-\zhalf}^{\zhalf}\rho(z)\dd z=0.5.
\label{eq:zhalf}
\end{equation}

The classical \citet{Too64} 2D criterion for gravitational instability
of razor-thin rotating discs is $Q<1$, where
 \begin{equation}
  Q\equiv\frac{\kappa\csiso}{\pi G\Sigma}
\label{eq:qtoomre}
 \end{equation}
 is Toomre's instability parameter at a given radius. Here $\kappa$ is
 the epicycle frequency, defined by
\begin{equation}  
\kappa^2 \equiv 4 \Omega^2+\frac{\dd \Omega^2}{\dd \ln R},
\end{equation}
with $\Omega$ is the angular frequency, assumed throughout this paper
to depend only on $R$, and $\csiso$ is the gas velocity dispersion
defined by $\csiso^2\equiv P/\rho$. The quantity $\sigma$ can be
interpreted also as the {\em isothermal} sound speed of the fluid, but
we prefer to refer to $\csiso$ generically as the velocity dispersion,
because this allows us to have more flexibility in the physical
interpretation of the fluid, which can represent not only a standard
smooth gas supported by thermal pressure, but also a turbulent gas, a
gas composed of discrete gas clouds or even a stellar distribution
(see Section \ref{sec:twocompdens}).

\subsection{The 3D gravitational instability criterion}
\label{sec:stabcrit}

We take as starting point the results of \citetalias{Nip23}, who
studied in 3D the local gravitational instability of rotating
stratified fluids.  Among the configurations considered by
\citetalias{Nip23}, we consider the simplest: vertically stratified
discs (section 3.3 of \citetalias{Nip23}), in which it is assumed
$\Omega=\Omega(R)$ and that the radial gradients of pressure and
density are locally negligible.  {\citetalias{Nip23} has shown that a
sufficient condition for local gravitational instability at given $(R,z)$
in such a vertically stratified disc is}
\begin{equation}
    \Qthreed\equiv\frac{\sqrt{\kappa^2+\nu^2}+\cs\hz^{-1}}{\sqrt{4\pi G\rho}}<1
    \qquad\mbox{(sufficient for instability)},
\label{eq:suff_instab_thick}  
\end{equation}
where $\nu$, defined by
\begin{equation}  
\nu^2\equiv \frac{\rhoz\pressz}{\rho^2}=\frac{\cs^2}{\gamma}\frac{\rhoz}{\rho}\frac{\pressz}{\press},
\label{eq:nusquared}
\end{equation}
is a frequency related to vertical pressure and density gradients
($\rhoz\equiv\de \rho/\de z$, $\pressz\equiv\de \press/\de z$), $\hz$
is a measure of the disc thickness, and $\cs=\sqrt{\gamma}\csiso$ is
the sound speed\footnote{The sound speed $\cs$ as defined here is in
general different from the gas velocity dispersion (or {\em
  isothermal} sound speed) $\csiso$. For instance, an ideal monoatomic
gas that undergoes adiabatic transformations has $\gamma=5/3$ and
$\cs=\sqrt{5/3}\csiso$, while for a fluid that undergoes isothermal
transformations $\gamma=1$ and $\cs=\csiso$. Note that in the adopted
notation $\gamma$ is not necessarily the heat capacity ratio: it is
the heat capacity ratio for an ideal gas undergoing adiabatic
transformations.}, with $\gamma\equiv(\rho/\press) \dd\press/\dd\rho$,
in which $\dd\press$ is the pressure variation corresponding to a
density variation $\dd \rho$ during a transformation.  Following
\citetalias{Nip23}, throughout the paper we assume $\hz=\hseventy$,
where $\hseventy$, defined by
\begin{equation}
\frac{1}{\Sigma}\int_{-\hseventy/2}^{\hseventy/2}\rho(z)\dd z=0.7,
\label{eq:hseventy}
\end{equation}
is the height of an infinitesimal-width strip centred on the midplane
containing 70\% of the mass per unit surface. {The scale height
  $\hz$ appearing in the definition of $\Qthreed$ derives from the
  approximation used by \citetalias{Nip23} for the linearized Poisson
  equation for radial perturbations. In
  Appendix~\ref{sec:pot_rad_pert} we assess quantitatively the
  validity of this approximation when $\hz=\hseventy$, by considering
  one-component discs. In the following we will adopt $\hz=\hseventy$
  also for two-component discs, assuming that it remains a
  sufficiently good approximation also in the presence of a second
  component.}

\subsection{Discs with self-gravitating isothermal vertical distribution}
\label{sec:sgidisc}

Before discussing more generic models, it is useful to recall the
properties of a disc in which the vertical density distribution is
that of the self-gravitating isothermal (SGI) slab \citep{Spi42}
\begin{equation}
\rho(z)=\rhozero\sech^2\ztilde,
\label{eq:rho_sgi}
\end{equation}  
where $\rhozero=\rho(0)$ is the density in the midplane and
$\ztilde\equiv z/b$, with $b\equiv \sigma/\sqrt{2\pi G\rhozero}$ a
scale height, where $\sigma$ is the $z$-independent velocity
dispersion. { The gravitational potential is
\begin{equation}
  \Phi(z)=-\csiso^2\ln \left[\frac{\rho(z)}{\rhozero}\right]
  \end{equation}
and the surface density is $\Sigma=2 b \rhozero$, so $\rhozero=\pi G
\Sigma^2/(2 \csiso^2)$.}  Using $\press(z)=\csiso^2\rho(z)$,
\Eq(\ref{eq:nusquared}) and \Eq(\ref{eq:rho_sgi}), we obtain that for
the SGI disc model we have
\begin{equation}
\nu^2=8\pi G \rhozero\tanh^2\ztilde.
\label{eq:nu_sgi}
\end{equation}
So, for a disc with angular velocity $\Omega=\Omega(R)$ and SGI
vertical distribution, at any given $R$ the sufficient condition for
instability (\ref{eq:suff_instab_thick}) becomes \citepalias{Nip23}
\begin{equation}
  \begin{split}
  \Qthreed=\sqrt{\frac{Q^2}{2}\cosh^2\ztilde+2\sinh^2\ztilde}
  +\sqrt{\frac{\gamma}{2}}\frac{b}{\hz}\cosh\ztilde<1\qquad\\
  \mbox{(sufficient
    for instability)}.
\label{eq:q3d_sgi}
\end{split}
\end{equation}

\section{Two-component discs with isothermal  vertical distributions}
\label{sec:twocompisoth}

We consider here two-component discs, in which each component has an
isothermal vertical density distribution.

\subsection{Equations}
\label{sec:twocompeq}

To describe a two-component disc, we use the index $i$
(for $i=1,2$) to label the generic $i$th component. At any given $R$
the vertical equilibrium is described by a two-component isothermal
slab (see \citealt{Ber22}) composed of two components with density
distributions $\rhoone(z)$ and $\rhotwo(z)$ in equilibrium in the
total gravitational potential \begin{equation}
  \Phi(z)=\Phione(z)+\Phitwo(z),
\label{eq:gravpot}
\end{equation}
where $\Phii$, satisfying 
\begin{equation}
 \frac{\dd^2 \Phii}{\dd z^2}
  =4\pi G\rhoi,
\label{eq:poisson_component}
\end{equation}
is the gravitational potential generated by $\rhoi$.
Each component is in vertical hydrostatic equilibrium in the total potential
$\Phi$, so 
\begin{equation}
  \frac{\dd \pressi}{\dd z}=-\rhoi\frac{\dd \Phi}{\dd z},
\label{eq:hydrostatic}
\end{equation}
where $\pressi(z)=\sigmai^2\rhoi(z)$, $\sigmai$ and $\rhoi(z)$ are,
respectively, the pressure, velocity dispersion and density of the
$i$th component.    Given that
$\sigmai$ is independent of $z$, \Eq(\ref{eq:hydrostatic}) can be
written as
\begin{equation}
\sigmai^2\frac{\dd \ln \rhoi}{\dd z}=-\frac{\dd \Phi}{\dd z},
\end{equation}
which has solution
\begin{equation}
\rhoi(z)=\rhozeroi e^{-\Phi/\sigmai^2},
\label{eq:rhoi}
\end{equation}
where $\rhozeroi$ is the midplane density of the $i$th component.

Combining \Eqs(\ref{eq:gravpot}), (\ref{eq:poisson_component}) and
(\ref{eq:rhoi}), we get the ordinary differential equation
\begin{equation}
  \frac{\dd^2 \Phi}{\dd z^2}
  =4\pi G\left(\rhozeroone  e^{-\Phi/\sigmaone^2}+\rhozerotwo  e^{-\Phi/\sigmatwo^2}\right),
\label{eq:d2phidz2}
\end{equation}
which, for given $\rhozeroone$, $\rhozerotwo$, $\sigmaone$ and $\sigmatwo$,
can be solved to obtain $\Phi(z)$, and then $\rhoone(z)$ and
$\rhotwo(z)$ from \Eq(\ref{eq:rhoi}).
\Eq(\ref{eq:d2phidz2}) can be written in dimensionless form as
\begin{equation}
\begin{split}
\frac{\dd^2 \Phitilde}{\dd \ztilde^2}
  =2\left(e^{-\Phitilde}+\xi e^{-\Phitilde/\mu}\right),
\end{split}
\label{eq:twocompdiffeq}
\end{equation}  
where $\Phitilde\equiv \Phi/\sigmaone^2$,
$\mu\equiv\sigmatwo^2/\sigmaone^2$,
$\xi\equiv\rhozerotwo/\rhozeroone$ and $\ztilde\equiv z/\bone$, with
$\bone\equiv\sigmaone/\sqrt{2\pi G\rhozeroone}$ a scale height.  The
two-component isothermal slab models as defined above have only two
free parameters: $\xi$ (the central density ratio of the two
components) and $\mu$ (their velocity dispersion ratio squared).
Assuming that the two components share the same $\kappa$, we define
the $Q$ parameter of the $i$th component as
\begin{equation}
\Qi=\frac{\kappa\sigmai}{\pi G \Sigmai},
\label{eq:qi}
\end{equation}
where $\Sigmai$ is the surface density of the $i$th component.

\subsection{Vertical density profiles}
\label{sec:twocompdens}

\begin{figure}
  \includegraphics[width=0.5\textwidth]{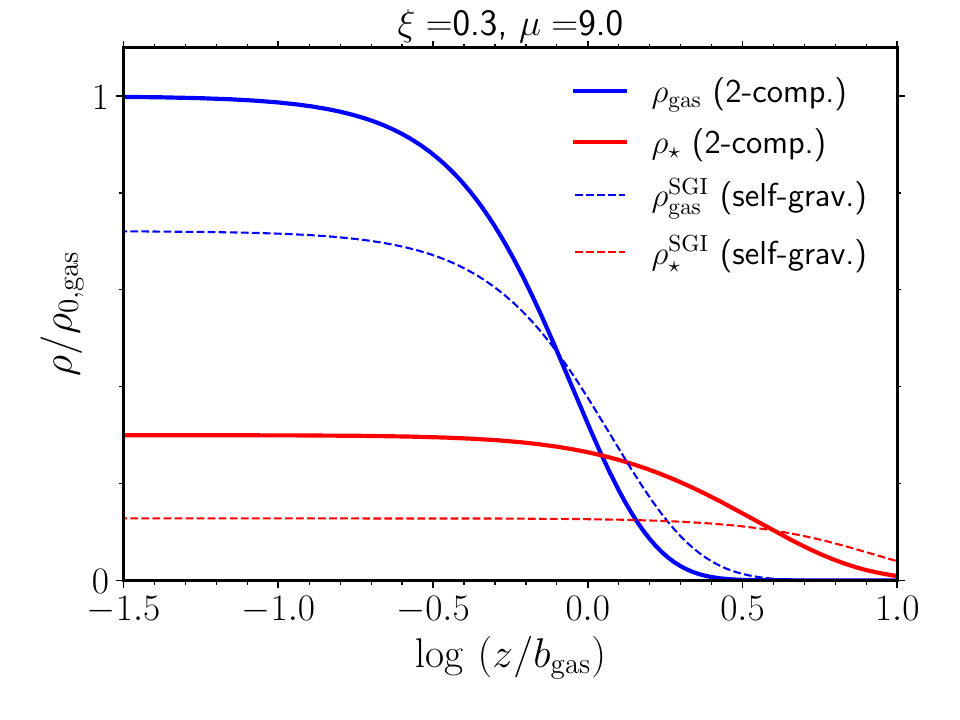}\\
  \includegraphics[width=0.5\textwidth]{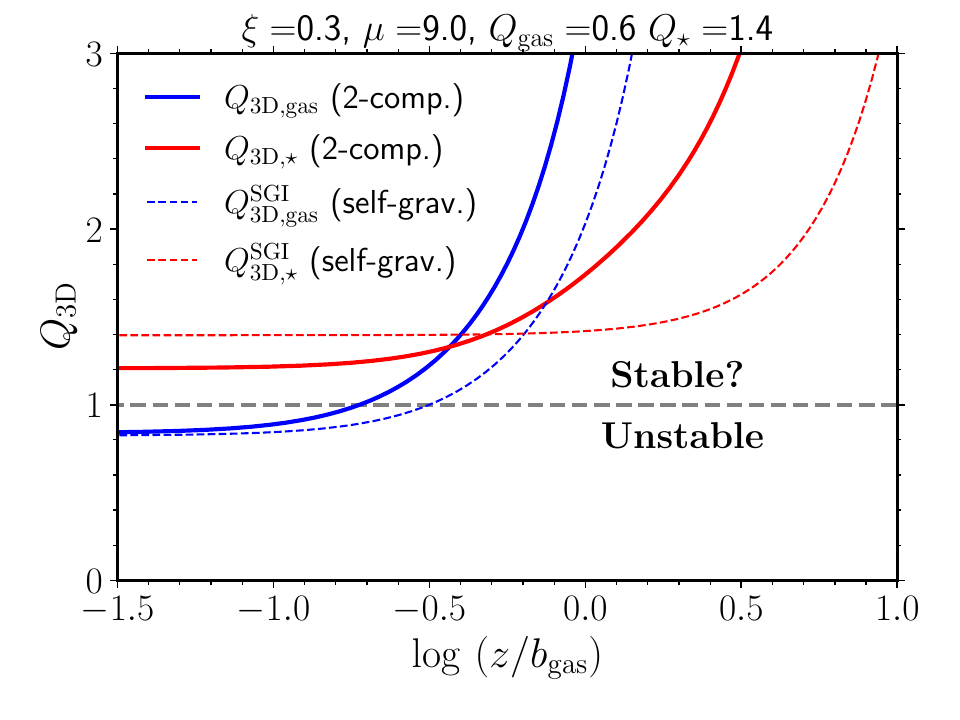}
  \caption{Upper panel. Vertical density profiles (solid curves) of
    gas and stars of a two-component isothermal disc with $\mu=9$
    and $\xi=0.3$. The dashed curves are the profiles of the
    corresponding SGI models with the same velocity dispersion and
    surface density.  Lower panel. Vertical $\Qthreed$ profiles of the
    same models as in the upper panel, assuming $\Qgas=0.6$ (and thus
    $\Qstar=1.4$). The horizontal dashed line indicates the
    instability threshold $\Qthreed=1$. Here $\rhozerogas$ and $\bgas$
    are, respectively, the midplane density and the scale height of
    the gaseous component of the two-component disc.}
    \label{fig:lowq}
\end{figure}

\begin{figure}
  \includegraphics[width=0.5\textwidth]{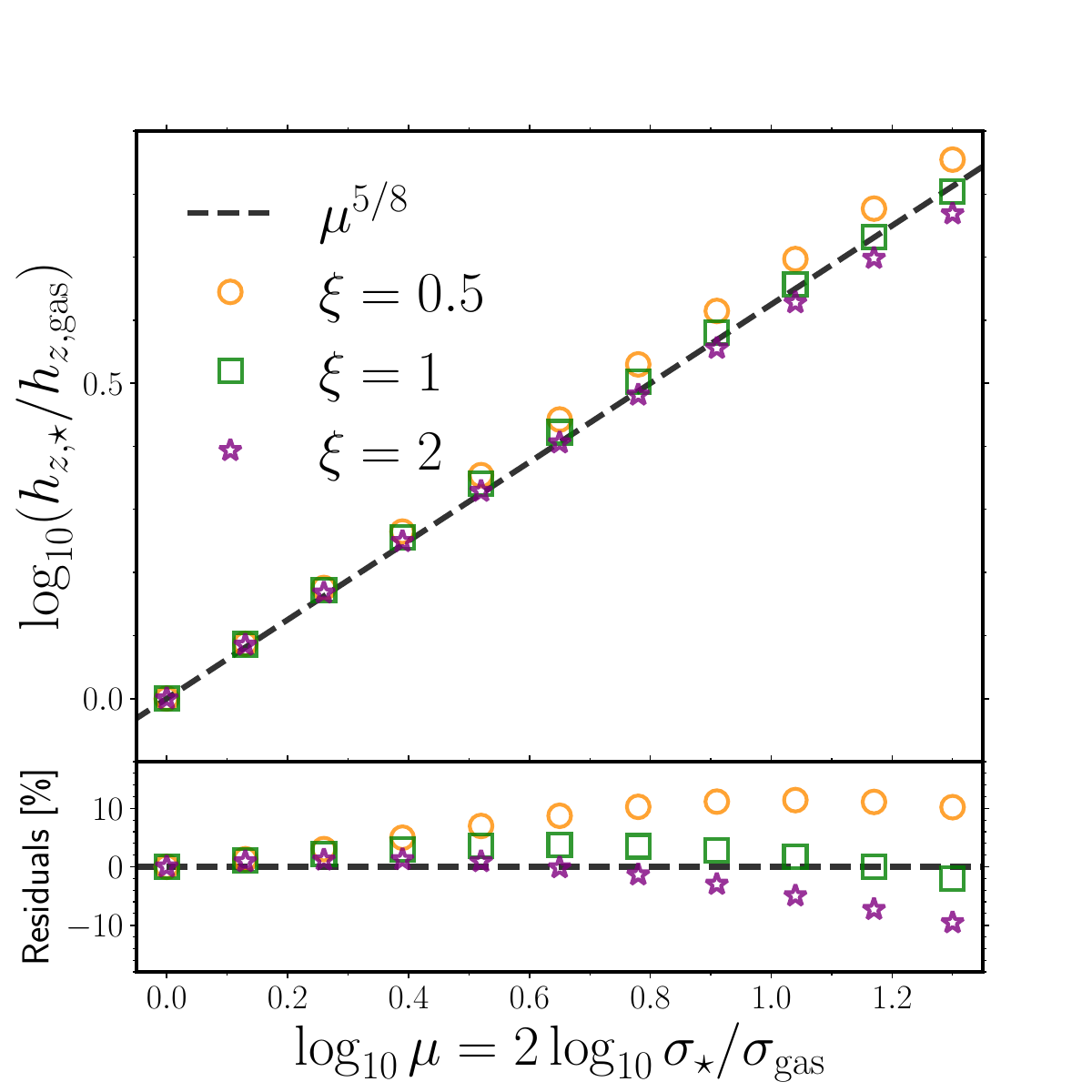}
  \caption{Upper panel. Stars-to-gas thickness ratio $\hzstar/\hzgas$
    as a function of the squared velocity-dispersion ratio $\mu$ for
    two-component isothermal slabs with different values of the
    midplane density ratio $\xi$ ($\hz=\hseventy$, defined by
    \Eq\ref{eq:hseventy}). Overplotted in grey is the power law
    $\mu^{5/8}$. Lower panel. Per cent residuals between the points
    and the power law of the upper panel.}
    \label{fig:ratios}
\end{figure}

\begin{figure}
      \includegraphics[width=0.47\textwidth]{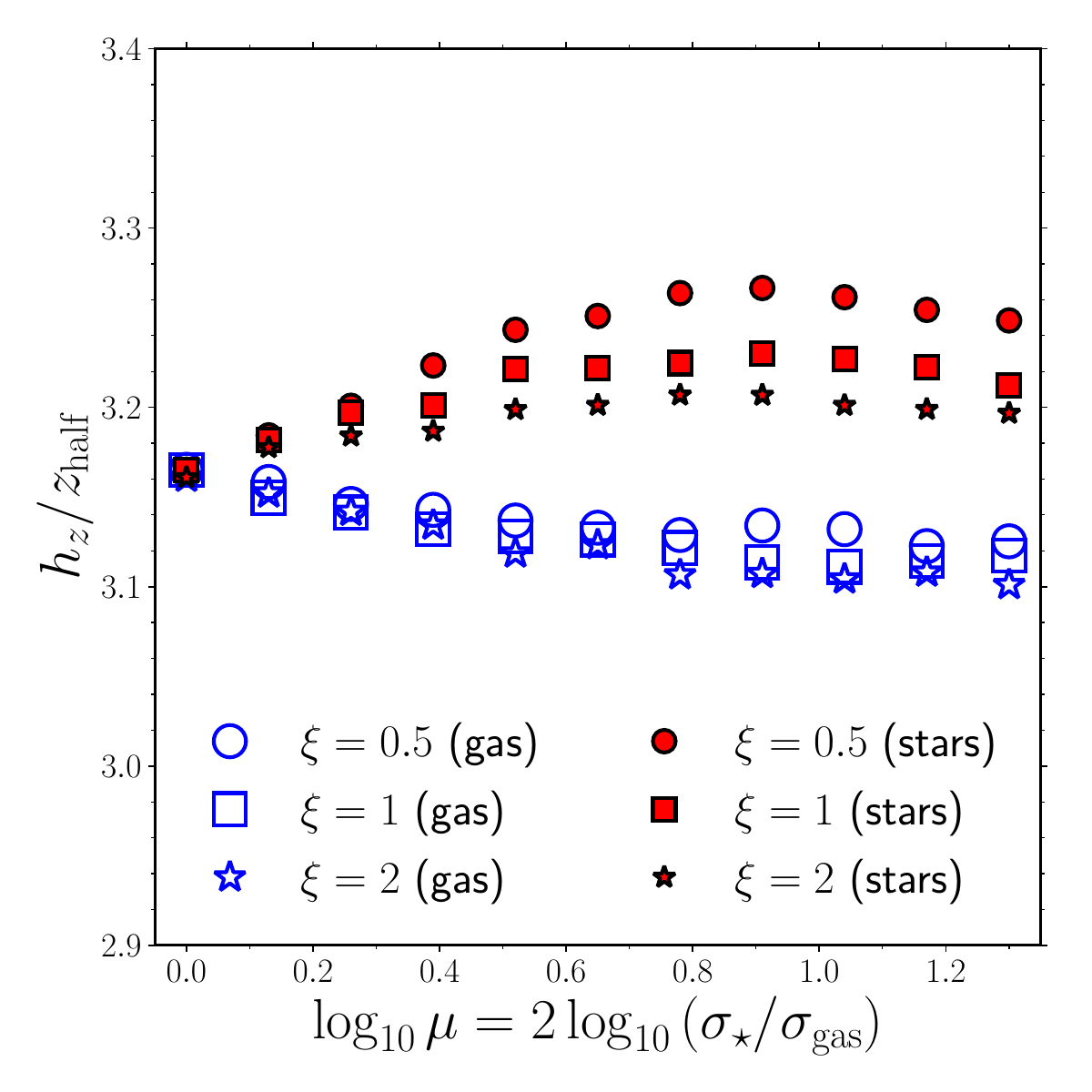}
  \caption{Ratios $\hzstar/\zhalfstar$ (filled symbols) and
    $\hzgas/\zhalfgas$ (empty symbols) as functions of the squared
    velocity-dispersion ratio $\mu$, for two-component isothermal
    slabs with different values of the midplane density ratio
    $\xi$. $\hz=\hseventy$ and $\zhalf$ are defined by
    \Eq(\ref{eq:hseventy}) and \Eq(\ref{eq:zhalf}), respectively.}
    \label{fig:ratios2}
\end{figure}

We present here an example of an equilibrium two-component isothermal
model obtained by solving numerically \Eq(\ref{eq:twocompdiffeq}) for
given values of the parameters $\mu$ and $\xi$, with boundary
conditions $\Phitilde=0$ and $\dd \Phitilde /\de \ztilde=0$ at
$\ztilde=0$. In particular, the numerical solutions were computed
using the {\tt Mathematica}\footnote{Wolfram Research, Inc.,
Mathematica, Version 14.0, Champaign, IL (2024).}  routine {\tt
  NDsolve}.

For the sake of definiteness, we will now refer to the specific case
of a galaxy with a gaseous disc and a stellar disc, thus we will use
the equations of Section~\ref{sec:twocompeq}, replacing the subscript
``1'' with ``gas'' and the subscript ``2'' with ``$\star$''.  In this
context, it is natural to assume that the effective gas pressure is
$\pressgas=\rhogas\sigmagas^2$, with $\sigmagas^2=\kB
T/\mbar+\sigmaturb^2$, where $T$ is the gas temperature, $\kB$ is the
Boltzmann constant, $\mbar$ is the mean molecular or atomic mass, and
$\sigmaturb$ is the turbulent velocity dispersion. This is motivated
by the fact that there is observational evidence that galactic gaseous
discs (either atomic or molecular) are turbulent and thus $\sigmaturb$
contributes substantially to $\sigmagas$ \citep[e.g.\ section 4.2.2
  of][]{CFN19,Bac20}.  Alternatively, the gas disc can be modelled as
a discrete distribution of gas clouds \citep[e.g.][]{Jef22}, in which
case $\sigmagas$ is just the cloud-cloud velocity dispersion.  The
vertical structure of a stellar disc is described by the same
equations used for a gas, but with $\sigmastar$ given by the vertical
stellar velocity dispersion.  Given that in a galaxy the stellar disc
is in general dynamically hotter than the gaseous disc \citep[at least
  for sufficiently old stellar populations;][]{van11}, without loss of
generality, we will limit ourselves to explore cases with
$\mu=\sigmastar^2/\sigmagas^2\geq 1$.

An example of the numerically computed vertical density distributions
of a two-component isothermal model with $\xi=0.3$ and $\mu=9$ is
given in the upper panel of Fig.\ \ref{fig:lowq}: for this model
$\Sigmastar/\Sigmagas\simeq 1.27$. For comparison, in the same diagram
we also plot the vertical gas density distributions of a gaseous SGI
slab with the same $\sigmagas$ and $\Sigmagas$ as the gaseous
component of the the two-component system and of a stellar SGI slab
with the same $\sigmastar$ and $\Sigmastar$ as the stellar component
of the the two-component system.  The stellar component, having higher
velocity dispersion, is more vertically extended than the gaseous
component.  Comparing the solid and dashed curves in the upper panel
of Fig.\ \ref{fig:lowq}, it is apparent how, for given surface density
and velocity dispersion, the volume density distribution depends on
the presence or absence of a second component.  Due to the deeper
gravitational potential, in the presence of the stellar disc, the
gaseous disc is more concentrated (i.e.\ with higher central volume
density) than in the self-gravitating case, and, vice versa, in the
presence of the gas disc, the stellar disc is more concentrated than
in the self-gravitating case.

For given vertical gravitational field the scale height increases for
increasing $\sigma$ \citep[see also][]{Bac19a}. Computing numerically
the ratios $\zhalfstar/\zhalfgas$ (see \Eq\ref{eq:zhalf}) and
$\hzstar/\hzgas$ (see \Eq \ref{eq:hseventy}), we find that they are
reasonably well described by the function $\mu^{5/8}$. This is
illustrated in Fig.\ \ref{fig:ratios}, where the ratio
$\hzstar/\hzgas$ is compared with $\mu^{5/8}$ for $\xi=0.5$, $\xi=1$
and $\xi=2$. A very similar behaviour is found for the scale-height
ratio, if one defines the scale height of the $i$th component as
$\Sigmai/(2\rhozeroi)$, as done in \citet[][who find a slope
  $3/5$]{Rom92} and \citet[][see their figure
  2]{Ber22}. Fig.\ \ref{fig:ratios2} shows how the ratios
$\hzstar/\zhalfstar$ and $\hzgas/\zhalfgas$ depend only weakly on
$\mu$ and $\xi$ in the explored range of values of these
parameters. For $\mu$ close to unity, $\hzstar/\zhalfstar\approx
\hzgas/\zhalfgas\approx 3.16$ independent of $\xi$. For large values
of $\mu$, $\hzstar/\zhalfstar\approx 3.22$ and
$\hzgas/\zhalfgas\approx 3.12$, with a weak dependence on $\xi$.

\subsection{Stability of each component assuming that the other component is unresponsive}
\label{sec:stabunresp}

{We study here the local gravitational instability of the
  two-component discs introduced in Section~\ref{sec:twocompdens},
  using the 3D instability criterion described in
  Section~\ref{sec:stabcrit}.} This is possible for each of the two
discs, under the simplifying assumption that the other disc is
unresponsive, i.e.\ it acts just as fixed external gravitational
potential and does not react dynamically to the perturbations (we
discuss the effects of a responsive disc in Section
\ref{sec:stabresp}). Though simplified, this approach is useful,
because, given that the back reaction of the other component favours
the instability \citep{Too64}, $\Qthreed<1$ is a sufficient condition
for instability also in the presence of a responsive second component.

{As in Section \ref{sec:twocompdens}, we interpret one of the two
  components as a stellar disc. Stellar discs are collisionless
  system, so, strictly speaking, their gravitational stability or
  instability cannot be assessed with the same methods used for fluid
  systems. However, it has been shown that treating the stellar disc
  as a fluid is a good approximation, at least in 2D stability
  analyses \citep[][]{Ber88,Raf01,Rom13}. To apply the criterion
  (\ref{eq:suff_instab_thick}) to the stellar component, we thus
  assume that the stellar component has isotropic velocity dispersion
  tensor and we model it as a fluid (see
  Section~\ref{sec:stellarfluid} for a discussion).}  Here we assume
that the gas and the stars undergo ``isothermal'' transformations,
i.e.\ such that $\dd \pressgas=\sigmagas^2\dd \rhogas$ and $\dd
\pressstar=\sigmastar^2\dd \rhostar$, which is in practice equivalent
to use $\gamma=1$ in the equations introduced in
Section~\ref{sec:prel}.  In principle we could consider full
axisymmetric disc models in which $\Sigmagas$, $\sigmagas$, $\kappa$
and thus $\Qgas=\kappa\sigmagas/(\pi G \Sigmagas)$, as well as the
corresponding stellar quantities, vary as functions of $R$
\citep[see][]{Bac24}.  Instead, as done in section 5 of
\citetalias{Nip23}, we exploit the mapping between $\Qgas$ and $R$ to
study the stability at given $R$, as a function of $z$, simply by
fixing a value of $\Qgas$.  Once $\Qgas$ is fixed, $\Qstar$ is
determined by the condition
$\Qstar/\Qgas=\sqrt{\mu}\Sigmagas/\Sigmastar$.

For the gaseous disc in the presence of stars, $\Qthreed$ can be computed
numerically once $\rhogas(z)$ and its derivative are computed, given
that $\pressgas=\sigmagas^2\rhogas$ (with $\sigmagas$ independent of $z$)
and that $\Qthreed$ (\Eq\ref{eq:suff_instab_thick}) can be rewritten for the gas component
as 
\begin{equation}
  \Qthreedgas(z)=
  \frac{1}{\sqrt{2\rhogastilde(z)}}
  \left[
    \sqrt{\Qgas^2\Sigmagastilde^2+\frac{1}{\rhogastilde^2(z)}\left(\frac{\dd \rhogastilde}{\dd \ztilde}\right)^2}
    +\frac{1}{\hzgastilde}
    \right].
  \label{eq:q3dgas}
\end{equation}
Similarly, for  the stellar component
\begin{equation}
  \Qthreedstar(z)=
  \frac{1}{\sqrt{2\rhostartilde(z)}}
  \left[
    \sqrt{\frac{\Qstar^2\Sigmastartilde^2}{\mu}+\frac{\mu}{\rhostartilde^2(z)}\left(\frac{\dd \rhostartilde}{\dd \ztilde}\right)^2}
    +\frac{\sqrt{\mu}}{\hzstartilde}
    \right],
  \label{eq:q3dstar}
\end{equation}
which can be obtained if $\rhostar(z)$ and its derivative are known
numerically.  In the above equations, where we have used $\gamma=1$,
$\rhogastilde=\rhogas/\rhozerogas$,
$\rhostartilde=\rhostar/\rhozerogas$, 
$\hzgastilde=\hzgas/\bgas$, $\hzstartilde=\hzstar/\bgas$,
$\Sigmagastilde\equiv \Sigmagas/(2\bgas\rhozerogas)$ and
$\Sigmastartilde\equiv \Sigmastar/(2\bgas\rhozerogas)$, where
$\bgas\equiv\sigmagas/\sqrt{2\pi G\rhozerogas}$.

Profiles of $\Qthreed$ for the gaseous and stellar components are
shown as solid curves in the lower panel of Fig.\ \ref{fig:lowq},
assuming $\Qgas=0.6$, for the same two-component disc models as in the
upper panel of Fig.\ \ref{fig:lowq}. For comparison, in the lower
panel of Fig.\ \ref{fig:lowq} we plot as dashed curves $\Qthreed$
(\Eq\ref{eq:q3d_sgi}) for the corresponding SGI gas and stellar discs
with the same surface density, velocity dispersion and $Q$.  This
example illustrates that the presence of a second component can affect
$\Qthreed$ in different ways: in this case, close to the midplane the
gas disc has higher $\Qthreedgas$ in the presence than in the absence
of the stellar disc (at fixed $\Qgas$, $\Sigmagas$ and $\sigmagas$),
while the stellar disc has lower $\Qthreedstar$ in the presence than
in the absence of the stellar disc (at fixed $\Qstar$, $\Sigmastar$
and $\sigmastar$).  This occurs because $\Qthreed$ depends not only on
the volume density (which, close to the midplane, is invariably higher
in the presence of a second component, for given $\Sigma$ and
$\sigma$), but also on the shape of the vertical density profile,
through $(\dd\rhoi/\dd z)/\rhoi$ and $\hzi$ (see \Eqs\ref{eq:q3dgas}
and \ref{eq:q3dstar}).

\begin{figure}
  \includegraphics[width=0.5\textwidth]{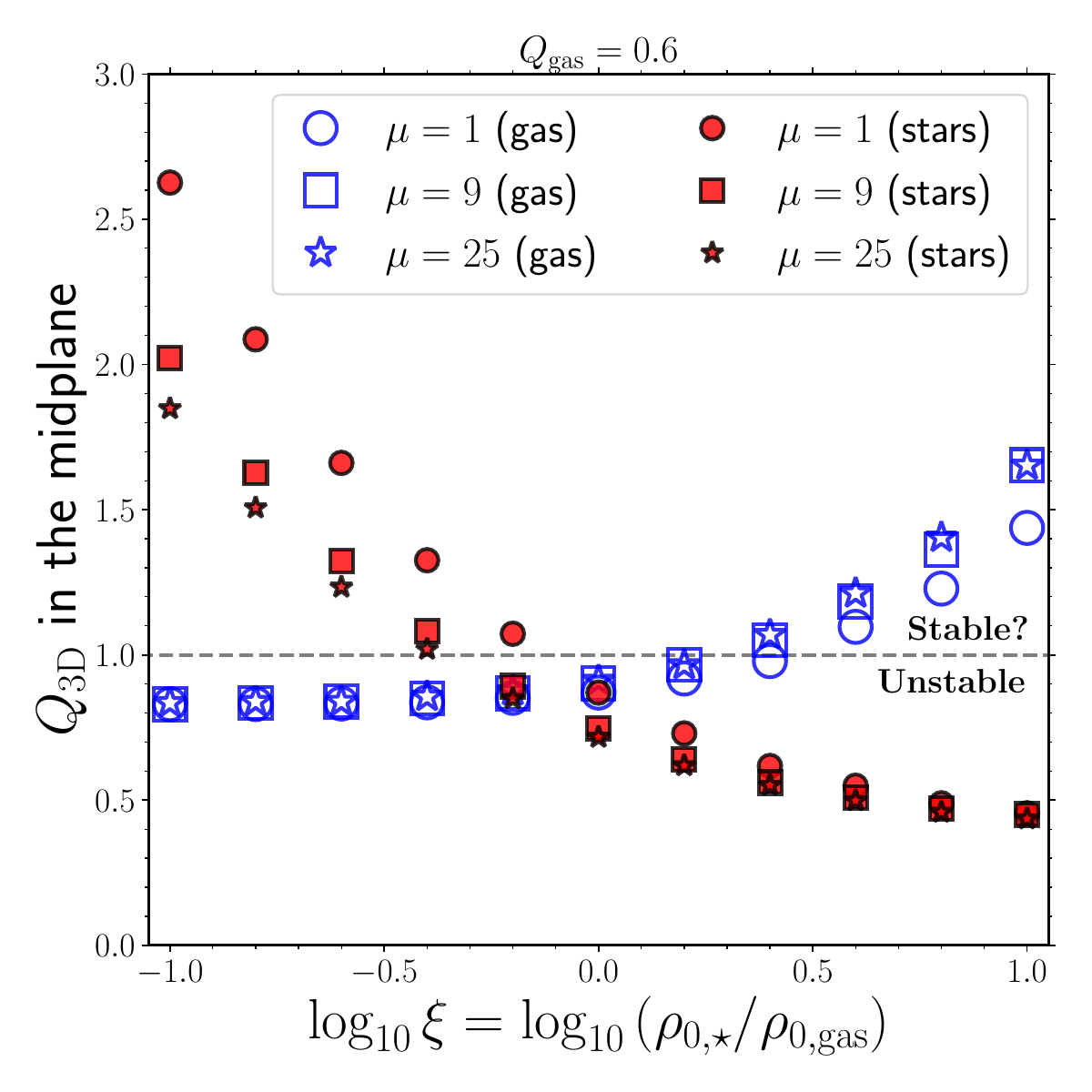}\\ \includegraphics[width=0.5\textwidth]{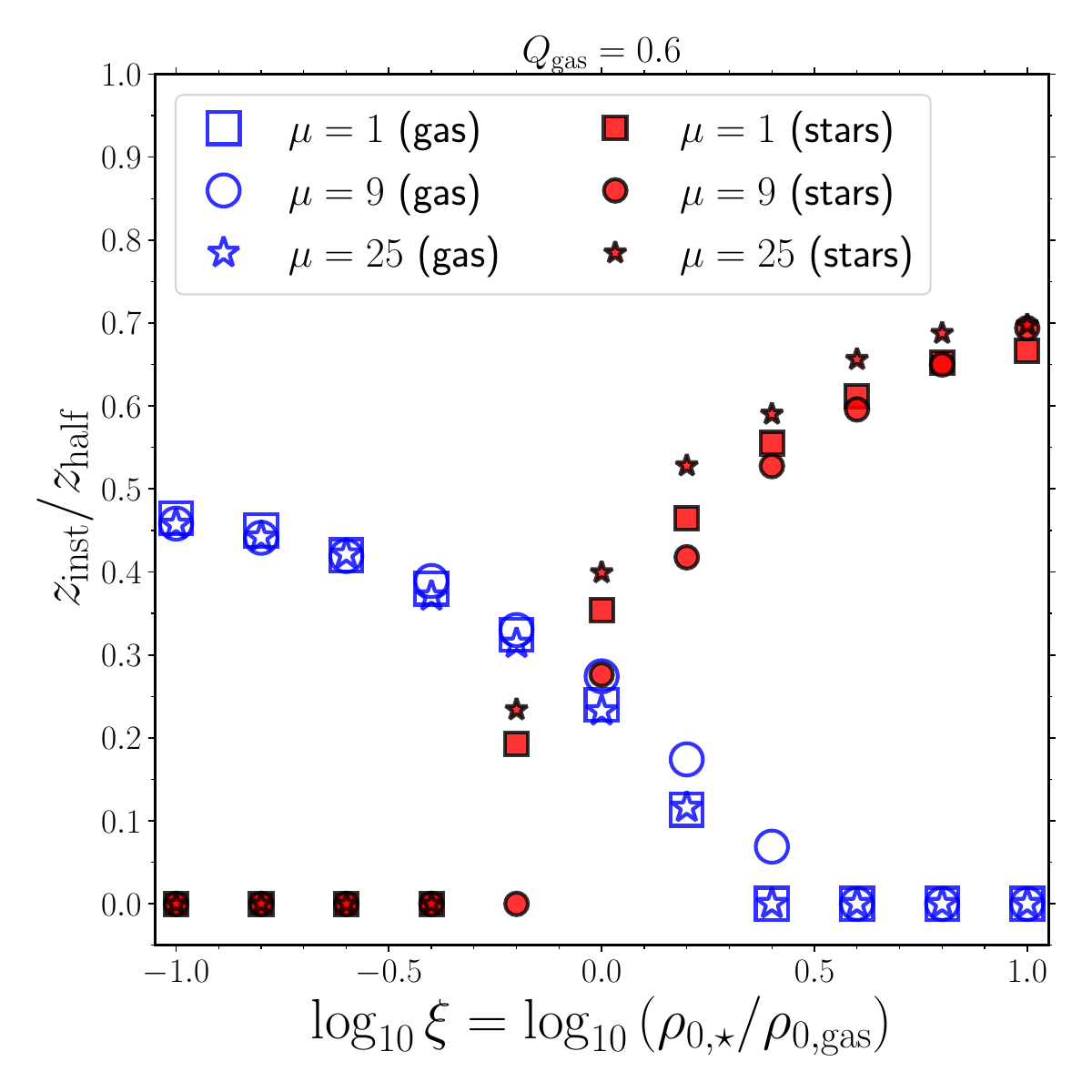}
  \caption{Upper panel. Midplane values of $\Qthreedstar$ (filled
    symbols) and $\Qthreedgas$ (empty symbols) as functions of the
    midplane star-to-gas density ratio $\xi$, for different values of
    the star-to-gas squared velocity dispersion ratio $\mu$.  The
    horizontal dashed line indicates the instability threshold
    $\Qthreed=1$. Lower panel. Extent of the unstable strip above the
    midplane for the stellar ($\zinststar$, normalized to
    $\zhalfstar$; filled symbols) and gaseous ($\zinstgas$, normalized
    to $\zhalfgas$; empty symbols) as functions of $\xi$ for different
    values of $\mu$. In both panels we assume $\Qgas=0.6$.}
    \label{fig:inst_xi}
\end{figure}

A systematic study of the family of two-component isothermal disc
models with $\Qgas=0.6$ for a range of values of $\mu$ and $\xi$ is
displayed as an illustrative example in Fig.\ \ref{fig:inst_xi}, which
shows that for the considered values of $\Qgas$ there is always
instability in the midplane, in the sense that either
$\Qthreedgas(0)<1$ or $\Qthreedstar(0)<1$, and that the behaviour of
the system depends weakly on $\mu$. In particular, from the upper
panel of Fig.\ \ref{fig:inst_xi}, plotting $\Qthreedgas(0)$ and
$\Qthreedstar(0)$ as functions of $\xi$, we see that when the gas disc
is dominant (low $\xi$) the instability is driven by the gas disc
($\Qthreedgas(0)<1$), while when the stellar disc is dominant (high
$\xi$) the instability is driven by the stars ($\Qthreedstar(0)<1)$.
The lower panel of Fig.\ \ref{fig:inst_xi} shows, for each component
in units of its $\zhalf$, the (above or below midplane) extent
$\zinst$ of the unstable region, defined by the condition
$\Qthreed(z)<1$ for $z<\zinst$ and $\Qthreed(z)>1$ for $|z|>\zinst$.
When one of the components is dominant ($\xi\ll 1$ or $\xi\gg 1$), we
have $0.5\lesssim \zinst/\zhalf\lesssim 0.7$ for $\Qgas=0.6$. When the
two components are comparable ($\xi\approx 1$) $\zinst/\zhalf\approx
0.3$ for $\Qgas=0.6$. {Fig.\ \ref{fig:inst_xi} suggests that the
  $\Qthreed$-based stability properties of two-component discs are
  essentially independent of $\mu$. At fixed $\Qgas$, the midplane
  density ratio $\xi$ determines which component drives the
  instability, but influences only weakly the overall instability
  properties of the system: the minimum between $\Qthreedgas(0)$ and
  $\Qthreedstar(0)$ varies only by about a factor of two when $\xi$
  spans two orders of magnitude (see upper panel of
  Fig.\ \ref{fig:inst_xi}). We recall, however, that the analysis
  based on $\Qthreed$ does not capture the full complexity of the
  problem of the instability of two-component discs, in which the
  coupling between stellar and gaseous perturbations plays an
  important role \citep[e.g.][]{Ber88,Rom13}. The effect of this
  coupling in 3D two-component discs is discussed in the next
  section.}

\section{Stability of two-component discs when both components are responsive}
\label{sec:stabresp}

In Section~\ref{sec:stabunresp} we studied the stability a 3D gaseous
disc in the presence of a stellar disc, but neglecting the
back-reaction of the stellar disc, and vice versa. Here we attempt to
address the more realistic, but more complicated question of the
instability of a 3D two-component disc when both components are
responsive.  {As done in Section \ref{sec:stabunresp}, we assume
  for simplicity that both components can be treated as fluids (even
  in the case in which one of the components is a stellar disc). The
  limitations of this assumption are discussed in Section
  \ref{sec:stellarfluid}).}

\subsection{Linear perturbation analysis}
\label{sec:linpert}

The governing equations are, for each component, the same as in
section 2.1 of \citetalias{Nip23}, but with the gravitational
potential $\Phi$ given by the sum of the gravitational potentials of
the two components (\Eq\ref{eq:gravpot}):
\begin{equation}\begin{split}
& \frac{\de \rhoi}{\de t} + \frac{1}{R}\frac{\de (R\rhoi \uRi)}{\de R} + \frac{\de (\rhoi \uzi)}{\de z} = 0 , \\
& \frac{\de \uRi}{\de t} +\uRi\frac{\de \uRi}{\de R}+\uzi\frac{\de \uRi}{\de z}  - \frac{\uphii^2}{R} = -\frac{1}{\rhoi}\frac{\de \pressi}{\de R} - \frac{\de \Phi}{\de R}- \frac{\de \Phiext}{\de R},\\
& \frac{\de \uphii}{\de t} +\uRi\frac{\de \uphii}{\de R}+\uzi\frac{\de \uphii}{\de z} + \frac{\uRi \uphii}{R}=0,\\
& \frac{\de \uzi}{\de t} +\uRi\frac{\de \uzi}{\de R}+\uzi\frac{\de \uzi}{\de z} = -\frac{1}{\rhoi}\frac{\de \pressi}{\de z} - \frac{\de \Phi}{\de z}- \frac{\de \Phiext}{\de z},\\
    & \frac{\pressi}{\gamma-1}\left(\frac{\de}{\de t} + \uvi\cdot\nabla\right) \ln (\pressi\rhoi^{-\gamma}) =0,\\
&\frac{1}{R}\frac{\de}{\de R}\left(R\frac{\de \Phii}{\de R}\right)+\frac{\de^2 \Phii}{\de z^2}=4\pi G\rhoi,
  \end{split}
\label{eq:axisymmeq}
\end{equation}
with $i=1,2$, where $\uvi=(\uRi,\uphii,\uzi)$ is the velocity of the
$i$th component, $\Phi=\Phi_1+\Phi_2$ and $\Phiext$ is any additional
gravitational potential (for instance that of the dark-matter halo)
assumed fixed\footnote{The back-reaction of a dark-matter halo is
supposed to be negligible, because within the disc its volume density
is expected to be significantly lower than the baryonic volume
density.}.

Here, for each component, we make the same assumptions described in
Section~\ref{sec:stabcrit}.  The unperturbed system is a stationary
rotating ($\uphii\neq0$) solution of \Eqs(\ref{eq:axisymmeq}) with no
meridional motions ($\uRi=\uzi=0$), locally negligible radial pressure
and density gradients, and $\uphione=\uphitwo=\Omega R$, with
$\Omega=\Omega(R)$. {We note that the fact that the radial
  pressure gradient is negligible in the unperturbed disc
  (i.e.\ $|\dd\pressi/\dd R|/\rhoi\ll\Omega^2 R$) implies that the
  epicyclic approximation condition $\sigmai\ll \kappa R$
  \citep[e.g.][]{Ber14} is satisfied. This can be seen by considering
  the following ordering:
\begin{equation}
  \frac{\sigmai^2}{R}\sim
  \sigmai^2 \left|\frac{\dd\ln\pressi}{\dd R}\right|\sim
  \frac{1}{\rhoi}\left|\frac{\dd\pressi}{\dd R}\right|\ll
  \Omega^2R\sim
  \kappa^2 R,
  \label{eq:ordering}
\end{equation}
where we have used $\left|{\dd\ln\pressi}/{\dd R}\right|\sim 1/R$ and
$\kappa\sim\Omega$.  }

As in \citetalias{Nip23}, we perturb the system (\ref{eq:axisymmeq})
writing a generic quantity $q=q(R,z,t)$ (such as $\rhoi$, $\pressi$,
$\Phii$ or any component of $\uvi$) as $q=\qunp+\delta q$, where the
(time independent) quantity $\qunp$ describes the stationary
unperturbed fluid and the (time dependent) quantity $\delta q$
describes the Eulerian perturbation. From now on, without risk of
ambiguity, we will indicate any unperturbed quantity $\qunp$ simply as
$q$. {Limiting ourselves to a linear stability analysis, we
  consider small ($|\delta q/q|\ll 1$) perturbations.  Given that the
  radial perturbations tend to be more unstable than the vertical ones
  \citep[][\citetalias{Nip23}]{Gol65a}, we consider purely radial
  disturbances with spatial and temporal dependence $\delta q
  \propto\exp[\im(\kR R-\omega t)]$, where $\omega$ is the frequency,
  and $\kR$ is the radial component of the wavevector, which we assume
  to be such that $|\kR|R\gg 1$, as it is standard in the
  short-wavelength approximation.}  Moreover, given that in the
one-component case $\Qthreed$ is lowest in the midplane, for
simplicity we focus only on the midplane, where we can
adopt\footnote{The order of the dispersion relation obtained from the
linear-perturbation analysis is higher when $\rhozi\neq 0$ and
$\presszi\neq 0$.}  $\rhozi=0$ and $\presszi=0$.  Under these
assumptions, the linearized perturbed equations are
\begin{equation}
  \begin{split}
    & - \im \omega \delta\rhoi+\im\kR\rhoi\delta\uRi=0,\\
    &-\im\omega\delta\uRi-2\Omega\delta\uphii=-\im\frac{\kR}{\rhoi}\delta\pressi-\im\kR\delta\Phione-\im\kR\delta\Phitwo,\\
  &-\im\omega\delta\uphii+\frac{\dd (\Omega R)}{\dd R}\delta\uRi+\Omega\delta\uRi=0,\\
    &-\im\omega\delta\uzi=0,\\
    &-\im\omega\frac{\delta\pressi}{\pressi}+\im\gamma\omega\frac{\delta\rhoi}{\rhoi}=0,\\
    & -\left(\kR^2+\hzi^{-2}\right)\delta\Phii=4\pi G\delta\rhoi,
  \end{split}
  \label{eq:twocomppertsys}
\end{equation}
with $i=1,2$, where the last equation is the perturbed Poisson
equation in the form of equation 14 of \citetalias{Nip23}. {As
  done in the previous sections, also here we assume $\hz=\hseventy$
  (\Eq\ref{eq:hseventy}; see Appendix~\ref{sec:pot_rad_pert} and
  Section~\ref{sec:stabcrit} for a discussion)}.

{The system (\ref{eq:twocomppertsys}) leads to the biquadratic dispersion relation
\begin{equation}
\omega^4-(\alphaone+\alphatwo)\omega^2+\alphaone\alphatwo-\betaone\betatwo=0,
    \label{eq:disprel}
\end{equation}
where we have defined
\begin{equation}
\alphaone\equiv B+s-A\frac{s}{s+\Eone},
\end{equation}
\begin{equation}
\alphatwo\equiv B+\mu s-\xi A\frac{\mu s}{\mu s+\Etwo},
\end{equation}
\begin{equation}
\betaone=A\frac{s}{s+\Eone}
\end{equation}
and
\begin{equation}
\betatwo=\xi A\frac{\mu s}{\mu s+\Etwo},
\end{equation}
with $B\equiv \kappa^2$, $s\equiv\gamma\csone^2\kR^2$, $A\equiv 4\pi
G\rhoone$, $\Eone\equiv\gamma\csone^2\hzone^{-2}$,
$\Etwo\equiv\gamma\cstwo^2\hztwo^{-2}$, $\mu \equiv\cstwo^2/\csone^2$ and
$\xi=\rhozerotwo/\rhozeroone$.}

The dispersion relation~(\ref{eq:disprel}) is formally identical to
that found by \citet{Jog84} in their 2D analysis, but with different
definitions and physical meaning of the coefficients, so our stability
analysis essentially follows that of section II of {\citet[][see
    also \citealt{Hof12} for further analysis of the same dispersion
    relation]{Jog84}}.  \Eq(\ref{eq:disprel}) has roots
\begin{equation}
\omega^2_{-,+} = \frac{1}{2} \left[
  (\alphaone+\alphatwo)\pm\sqrt{(\alphaone+\alphatwo)^2-4(\alphaone\alphatwo-\betaone\betatwo)}
  \right].
\end{equation}
It is straightforward to show that the argument of the square root is
always positive, so the solutions are always real. Given that
$\omega_-^2\leq \omega_+^2$, we can focus on the root $\omega_-^2$ to
study the stability. When either $\alphaone<0$ or $\alphatwo<0$,
$\omega_-^2<0$, thus we have instability. We note that the condition $\alphaone<0$ is equivalent to 
\begin{equation}
\Fone(\kR)\equiv\frac{A s }{(s+B)(s+\Eone)}>1,
\label{eq:Fone}
\end{equation}
which, following the analysis reported in appendix B3 of
\citetalias{Nip23}, leads to the sufficient condition for instability
\begin{equation}
\frac{\sqrt{B}+\sqrt{\Eone}}{\sqrt{A}}<1,
\label{eq:resp_suff_cond_xi0}
  \end{equation}
which is just $\Qthreedone<1$ in the special case $\rhozone=0$
considered in this section. It is straightforward to show that
$\alphatwo<0$, i.e.
\begin{equation}
\Ftwo(\kR)=\frac{\xi A \mu s }{(s+B)(\mu s+\Etwo)}>1,
\label{eq:Ftwo}
\end{equation}
leads to
\begin{equation}
\frac{\sqrt{B}+\sqrt{\Etwo}}{\sqrt{A}}<1,
\label{eq:resp_suff_cond_xi0_2}
\end{equation}
that is  $\Qthreedtwo<1$ when $\rhoztwo=0$.  Thus, as
expected, when one of the components has $\Qthreed<1$ in the midplane
(i.e.\ it would be unstable, neglecting the back-reaction of the other
component), the two-component system results unstable also when the
back-reaction is accounted for.

When both $\alphaone>0$ (i.e.\ $\Fone<1$) and $\alphatwo>0$
(i.e.\ $\Ftwo<1$), the condition to have $\omega_-^2<0$ (and thus
instability) is
\begin{equation}
\alphaone\alphatwo-\betaone\betatwo<0,
\end{equation}
i.e.\
\begin{equation}
  \Foneplustwo(\kR)=\Fone(\kR)+\Ftwo(\kR)>1,
\label{eq:Foneplustwo}
\end{equation}
where the destabilizing effect of the second responsive component is
apparent.  This is a manifestation of the fact that, because of the
mutual back-reactions of the two components, a two-component system can
be gravitationally unstable even when each component would be stable
if taken individually.

When $\Qthreedone>1$ and $\Qthreedtwo>1$, and thus $\Fone<1$ and
$\Ftwo<1$ for all $\kR$, $\Foneplustwo>1$ is a sufficient condition
for instability.

\begin{figure}
  \includegraphics[width=0.5\textwidth]{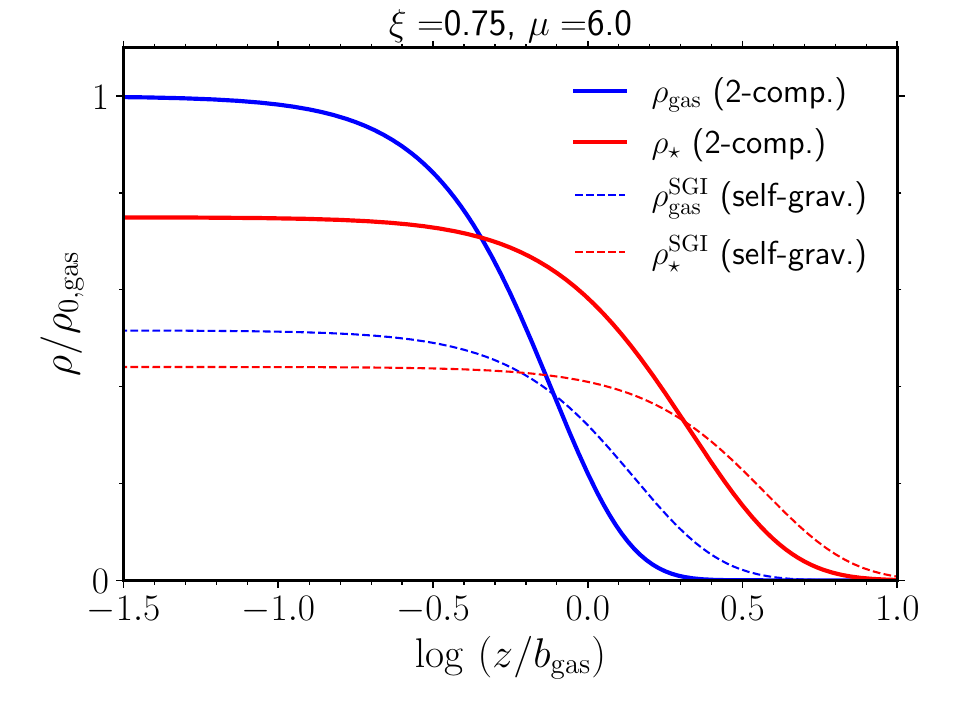}\\
  \includegraphics[width=0.5\textwidth]{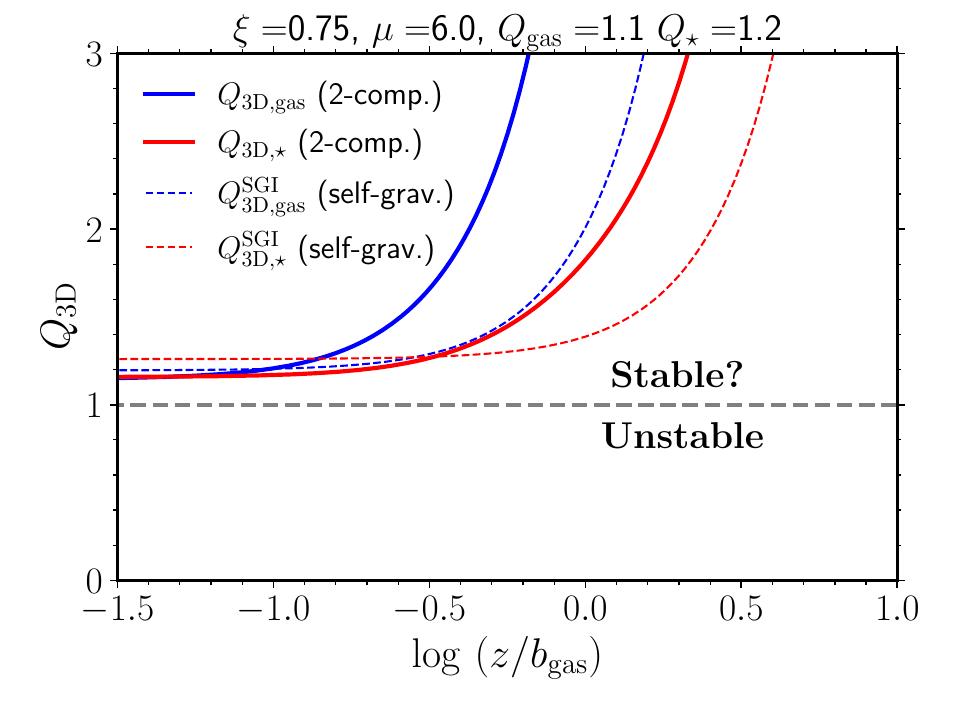}\\
  \caption{Same as Fig.\ \ref{fig:lowq}, but for a two-component
    isothermal disc model with $\xi=0.75$, $\mu=6$, $\Qgas=1.1$ and
    $\Qstar=1.2$.  }
    \label{fig:highq}
\end{figure}
\begin{figure}
  \includegraphics[width=0.5\textwidth]{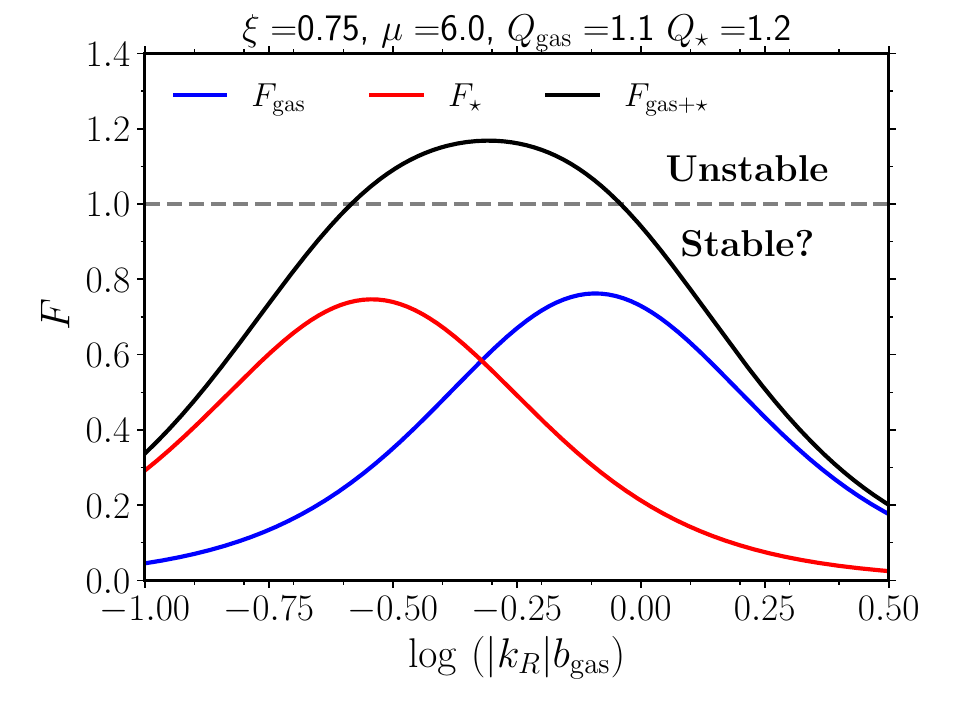}
  \caption{The functions $\Fgas(\kR)$, $\Fstar(\kR)$ and
    $\Fgasplusstar(\kR)$ (see \Eqs\ref{eq:Fgas}, \ref{eq:Fstar} and
    \ref{eq:Fgasplusstar}) for the same model as in
    Fig.\ \ref{fig:highq}. {The unstable perturbations are those
      with $\Fgasplusstar>1$.}}
    \label{fig:Fgasplusstar}
\end{figure}

\subsection{Application to two-component discs with stellar and gaseous components}
\label{sec:fgasplustar}

In order to apply quantitatively the results of
Section~\ref{sec:linpert}, let us specialize to the case of a
two-component disc with gas and stellar components. Relabeling now the
component indices $1$ and $2$ as ``gas'' and ``$\star$'',
respectively, as done in Sections \ref{sec:twocompdens} and
\ref{sec:stabunresp}, it follows from the calculations of
Section~\ref{sec:linpert} that when either $\Qthreedgas<1$ or
$\Qthreedstar<1$ in the midplane, we have instability.  {Let us thus
focus on the case in which in the midplane $\Qthreedgas>1$ and
$\Qthreedstar>1$, so that for all $\kR$ we have $\alphagas>0$ and
$\alphastar>0$ , which, using the definitions (\ref{eq:Fone}) and
(\ref{eq:Ftwo}) can be written, respectively, as
\begin{equation}
  \Fgas(\kR)= \frac{2\kRtilde^2 }{(\gamma\kRtilde^2+\Qgas^2\Sigmagastilde^2)(\kRtilde^2+\hzgastilde^{-2})}<1
  \label{eq:Fgas}
\end{equation}
and
\begin{equation}
\Fstar(\kR)= \frac{2
  \xi \kRtilde^2}{(\mu \gamma\kRtilde^2+\Qstar^2\Sigmastartilde^2/\mu)(\kRtilde^2+\hzstartilde^{-2})}<1,
\label{eq:Fstar}
\end{equation}
where $\kRtilde\equiv\kR\bgas$.  The condition for a disturbance with
radial wavenumber $\kR$ to be unstable is \Eq(\ref{eq:Foneplustwo}),
which in this case reads
\begin{equation}
  \Fgasplusstar(\kR)\equiv\Fgas(\kR)+\Fstar(\kR)>1.
  \label{eq:Fgasplusstar}
\end{equation}
}

We then consider two-component isothermal discs in which the vertical
distributions are computed numerically as described in Section
\ref{sec:twocompdens}.  {In the example shown in
  Fig.s\ \ref{fig:highq} and \ref{fig:Fgasplusstar}, where we assume
  $\gamma=1$ as done in Section \ref{sec:stabunresp}, the values of
  the parameters ($\xi=0.75$, $\mu=8$, $\Qgas=1.1$ and $\Qstar=1.2$)
  are such that the two-component isothermal disc would not be considered
  unstable according to the analysis of Section \ref{sec:stabunresp},
  but turns out to be unstable when the back-reaction of one component
  onto the other is taken into account.} Similar to the model shown in
Fig.\ \ref{fig:lowq}, the model shown in Fig.\ \ref{fig:highq}, for
which $\Sigmastar/\Sigmagas\simeq 2.26$, has higher gas than stellar
density close to the midplane (see upper panel of
Fig.\ \ref{fig:highq}), but $\kappa$ is assumed to be such that
$\Qthreedgas>1$ and $\Qthreedstar>1$ (see lower panel of
Fig.\ \ref{fig:highq}). Nevertheless, the system is gravitationally
unstable, because there is a range of values of the wavenumber such
that $\Fgasplusstar>1$ (Fig.\ \ref{fig:Fgasplusstar}).

We stress that the criterion (\ref{eq:Fgasplusstar}) can be applied to
observed discs, because all the coefficients of the functions $\Fgas$
and $\Fstar$ can be estimated from observable quantities. In
particular, to fully determine the function $\Fgasplusstar(\kR)$ at
given $R$ one needs estimates of $\Sigmagas$, $\Sigmastar$,
$\rhozerogas$, $\rhozerostar$, $\hzgas$, $\hzstar$, $\sigmagas$,
$\sigmastar$, and $\kappa$, which can all be inferred from
observational data \citep[see][]{Bac24}.  {Then, to draw
  conclusions on the local gravitational instability of the system at
  given $R$ in the disc midplane, it is sufficient to evaluate
  $\Fgasplusstar$ over a range of values of $|\kR|$, with lower limit
  between $1/R$ and $1/\hz$ (see Section~\ref{sec:linpert}) and upper
  limit $\gg 1/\hz$ (because the unstable disturbances have $|\kR|\hz$
  of the order of unity; \citealt{Gol65a}; \citetalias{Nip23}).}

\subsection{Limitations of the application to stellar discs}
\label{sec:stellarfluid}

{        
In Sections \ref{sec:stabunresp}, \ref{sec:linpert} and
\ref{sec:fgasplustar}, we assumed that the component interpreted as a
stellar disc has isotropic velocity dispersion tensor and we treated
it as a fluid in the perturbation analysis. This approach has some
limitations, which we discuss here.  In the case of 2D disc models,
comparisons between fluid and kinetic stability analyses
\citep[e.g.][]{Raf01} have shown that modelling a stellar disc as a
fluid turns out to be a sufficiently good approximation, provided the
quantity entering the disc gravitational instability diagnostics are
correctly interpreted in terms of the collisionless quantities.  For
instance, it is the radial stellar velocity dispersion $\sigmastarR$
that enters the definition of $\Qstar$. Extending the same line of
reasoning to our 3D analysis, when assuming that $\Qthreed$ is an
approximate instability indicator for collisionless discs, we have to
take care of relating the fluid quantities appearing in $\Qthreed$ to
collisionless quantities. As it is clear from the analysis of
\citetalias{Nip23} (see also \Eqs\ref{eq:axisymmeq} and
\ref{eq:twocomppertsys}), the sound speed $\cs=\gamma\sigma$ appearing
in \Eq(\ref{eq:suff_instab_thick}) derives from a radial derivative of
the pressure, so in this case $\sigma$ must be identified with
$\sigmastarR$. The quantity $\cs^2/\gamma=\sigma^2$ appearing in
$\nu^2$ (see \Eq\ref{eq:nusquared}) derives from a vertical derivative
of the pressure, so in this case $\sigma$ must be identified with the
vertical velocity dispersion $\sigmastarz$. Finally, the scale height
$\hz$, for given midplane density, is expected to increase for
increasing $\sigmastarz$, while being independent of $\sigmastarR$.

Focusing for simplicity on the midplane (where we expect $\nu\approx 0$) and assuming $\gamma=1$, 
we can thus rewrite $\Qthreed$ (\Eq\ref{eq:suff_instab_thick}) for a
stellar disc as
\begin{equation}
  \Qthreedstar\approx
  \sqrt{\frac{\kappa^2}{4\pi G\rhozerostar}}
    +\frac{\sigmastarR}{\sigmastarz}\left(\frac{\hz}{\sigmastarz/\sqrt{4\pi G\rhozerostar}}\right)^{-1}.
\label{eq:q3dstar}  
\end{equation}
Given that neither the first term in the r.h.s.\ nor the quantity in
parentheses depend on the velocity ellipsoid, this equation shows
that, for instance, a stellar disc with $\sigmastarR>\sigmastarz$ has
in fact higher $\Qthreedstar$ (and thus is more stable) than estimated
under the assumption of isotropic ($\sigmastarR/\sigmastarz=1$) velocity
distribution. Similar considerations apply to the instability
indicators considered in Sections~\ref{sec:linpert} and
\ref{sec:fgasplustar}, given the relationship between $\Qthreedstar$
and $\Fstar$.

There is observational evidence \citep{Ger12,Mar17,Pin18,Mac19,Mog19}
that in present-day stellar discs the velocity dispersion tensor is in
general anisotropic, with typically $\sigmastarR>\sigmastarz$, a
finding which is supported also by theoretical models
\citep{Rod13,Wal21}. The velocity ellipsoid of stars in
higher-redshift disc is hard to measure observationally, but it is
reasonable to expect that higher-redshift discs are dynamically
younger and thus more isotropic, having inherited the velocity
distribution from the gas from which their stars have formed. This
picture is supported by the finding that in the Milky Way the vertical
scale height of the youngest stellar population is similar to that of
the cold gas disc \citep[e.g.][]{Bac19b}
}

\section{One-component self-gravitating discs with  polytropic vertical distributions}
\label{sec:poly}

The vertical structure of discs is often modelled as isothermal, but
in fact the velocity dispersion can in general have a vertical
gradient.  Here we explore how the stability properties of a stratified
disc depend on such gradient, by exploring one-component
self-gravitating discs with polytropic vertical distributions.

\subsection{Equations}
\label{sec:poly_eq}

For a self-gravitating disc with polytropic vertical distribution, at
given $R$
\begin{equation}
  \press(z)=\presszero\left[\frac{\rho(z)}{\rhozero}\right]^{\gammap},
\label{eq:ploytropic}
\end{equation}
where $\gammap\equiv 1+1/n$ is the polytropic exponent, $n$ is the
polytropic index, $\rhozero=\rho(0)$ and $\presszero=\press(0)$. We
assume that the gas is in vertical hydrostatic equilibrium in its own
gravitational potential, so the vertical stratification is the same as
that of a polytropic slab (chapter 2 of \citealt{Hor04}; see also
\citealt{Iba84}). \Eq(\ref{eq:ploytropic}), combined with the
hydrostatic equilibrium equation
\begin{equation}
  \frac{\dd \press}{\dd z}=-\rho\frac{\dd \Phi}{\dd z},
\label{eq:hydrostatic2}
\end{equation}
and with the Poisson equation
\begin{equation}
\frac{\dd^2 \Phi}{\dd z^2}  =4\pi G\rho,
\label{eq:poisson}
\end{equation}
gives for finite $n$ the Lane-Emden equation
\begin{equation}
\frac{\dd^2 \theta}{\dd \zeta^2}=\pm\theta^n,
\label{eq:lanemden}
\end{equation}
where $\theta\equiv (\rho/\rhozero)^{1/n}$ and
\begin{equation}
\zeta\equiv z/\left[(n+1)\presszero/(4\pi G \rhozero^2)\right]^{1/2}.
\end{equation}
In the right hand side of \Eq(\ref{eq:lanemden}) the sign is minus
when $-1<n<\infty$ (i.e.\ $\gammap<0\,\cup\,\gammap>1$), while it is
plus when $-\infty<n<-1$ (i.e.\ $0<\gammap<1$).  When $n=\pm\infty$
($\gammap=1$), combining \Eqs(\ref{eq:ploytropic}-\ref{eq:poisson}) we
get
\begin{equation}
\frac{\dd^2 \theta}{\dd \zeta^2}=e^{-\theta},
\label{eq:lanemdenisoth}
\end{equation}
whose analytic solution is the SGI slab described in Section
\ref{sec:sgidisc}.  We assume that the fluid undergoes barotropic
transformations of the form $\press\propto\rho^\gamma$, so
$\dd\press=\gamma(\press/\rho)\dd\rho$ and $\cs^2=\gamma\csiso^2$.  We
limit ourselves to convectively stable polytropic distributions,
i.e.\ satisfying the Schwarzschild stability criterion
$\gammap\leq\gamma$. In particular, in this section we consider two
representative specific cases.
\begin{enumerate}
  \item A monoatomic ideal gas undergoing adiabatic transformations
    with $\gamma=5/3$, for which $\cs=\sqrt{5/3}\csiso$ and
    convectively stable distributions are those with $\gammap\leq
    5/3$, (i.e.\ $n<0\,\cup\,n\geq 3/2$).
\item A fluid undergoing isothermal transformations with $\gamma=1$,
  for which $\cs=\csiso$ and convectively
stable distributions are those with $\gammap\leq1$ ($n\leq0$).
\end{enumerate}
We note however that for $\gammap<0$ ($-1<n<0$) the density increases
for increasing distance from the midplane \citep{Via74}, so in all
cases we exclude from our analysis $\gammap<0$ ($-1<n<0$).

\subsection{Vertical density and velocity-dispersion profiles}

Given that we focus on the cases $\gamma=5/3$ and $\gamma=1$, and that
we require convective stability (see Section~\ref{sec:poly_eq}), we
present here polytropic distributions with polytropic exponent in the
range $0< \gammap\leq 5/3$ (i.e.\ $n\geq 3/2\,\cup\,n<-1$).

\begin{figure}
  \includegraphics[width=0.5\textwidth]{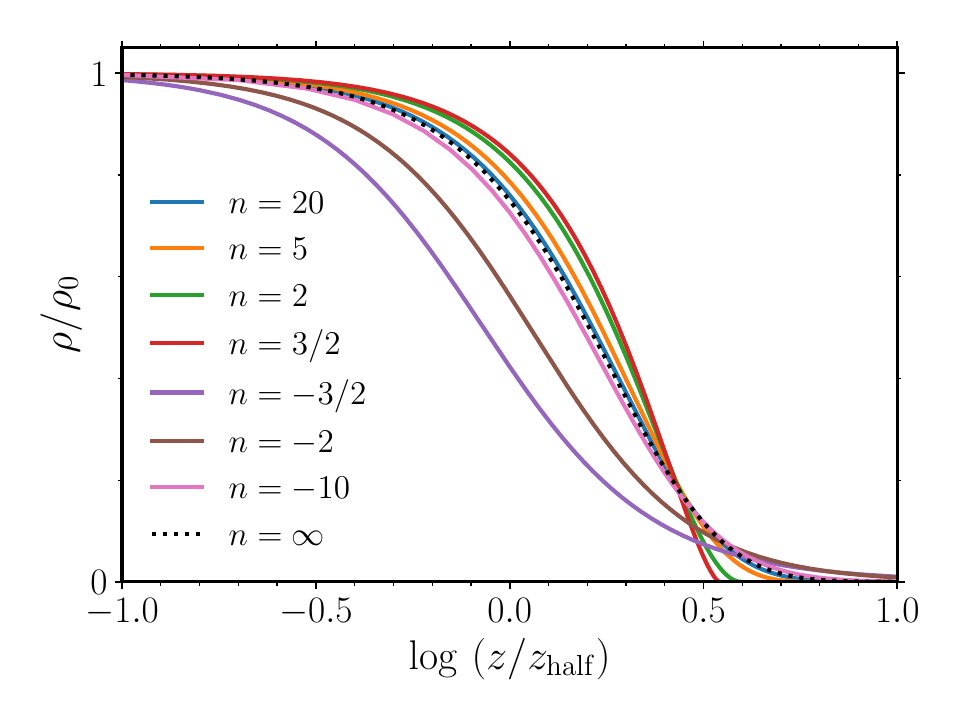}\\
  \includegraphics[width=0.5\textwidth]{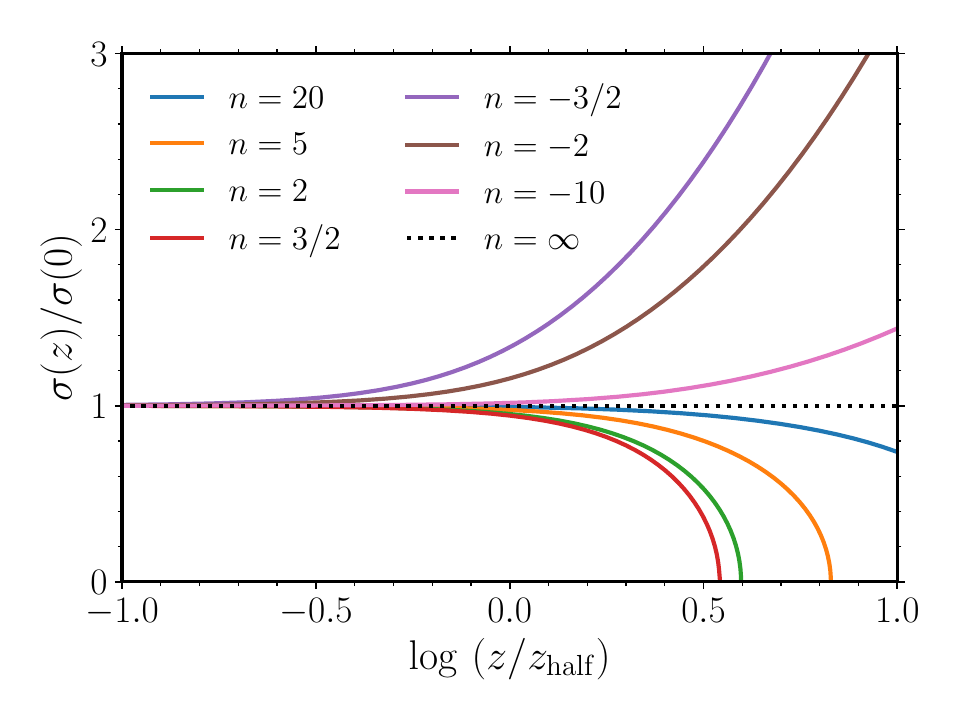}\\
  \caption{{Vertical density (upper panel) and velocity-dispersion
      (lower panel) profiles (solid curves) of one-component
      self-gravitating slabs with polytropic vertical distributions
      for different values of the polytropic index $n$.} The case
    $n=\infty$ (dotted curves) is the analytic SGI vertical
    distribution.}
    \label{fig:poly_rho}
\end{figure}

We solved numerically \Eq\ref{eq:lanemden} with boundary conditions
$\theta=0$ and $\dd \theta /\de \zeta=0$ at $\zeta=0$ using the {\tt
  Mathematica} routine {\tt NDsolve}.  We find it convenient to
normalize the vertical coordinate $z$ to the half-mass half-height
$\zhalf$.  The vertical density and velocity-dispersion profiles are
shown in Fig.\ \ref{fig:poly_rho}. Note that the velocity dispersion
$\sigma$ is defined by $\sigma^2=\press/\rho$, so the profiles in the
lower panel of Fig.\ \ref{fig:poly_rho} can also be interpreted as
profiles of $\sqrt{T}$ for a disc vertically supported by thermal
pressure. We verified that the profiles shown in
Fig.\ \ref{fig:poly_rho} are consistent with those tabulated by
\citet{Iba84} and \citet{Hor04} for the values of $n$ that we have in
common.

\subsection{Stability}
\label{sec:polstab}

\begin{figure}
  \includegraphics[width=0.5\textwidth]{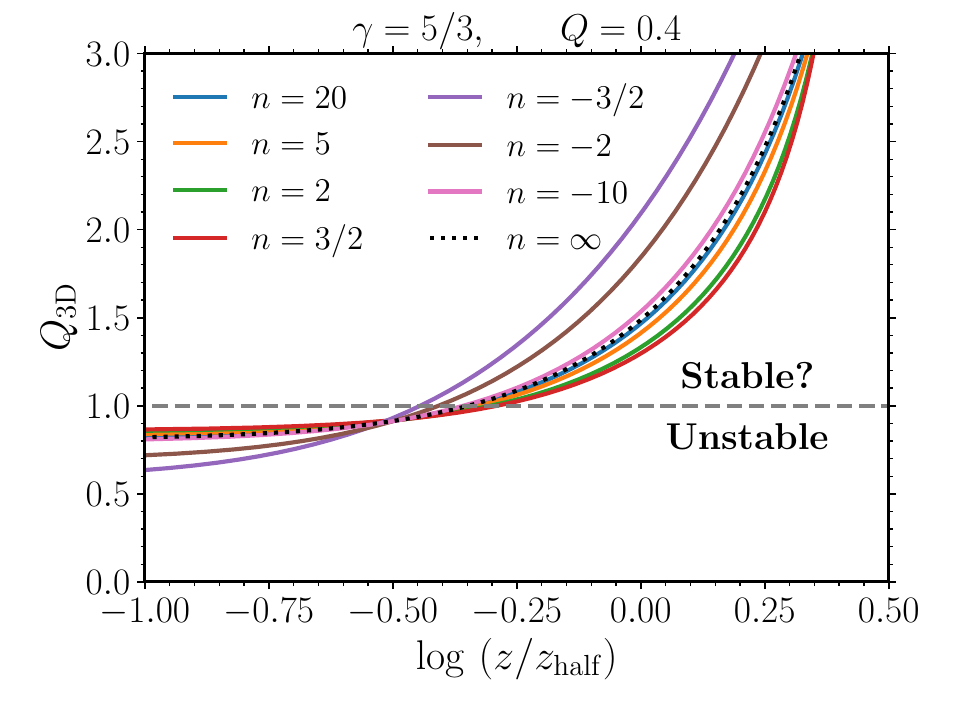}\\
    \includegraphics[width=0.5\textwidth]{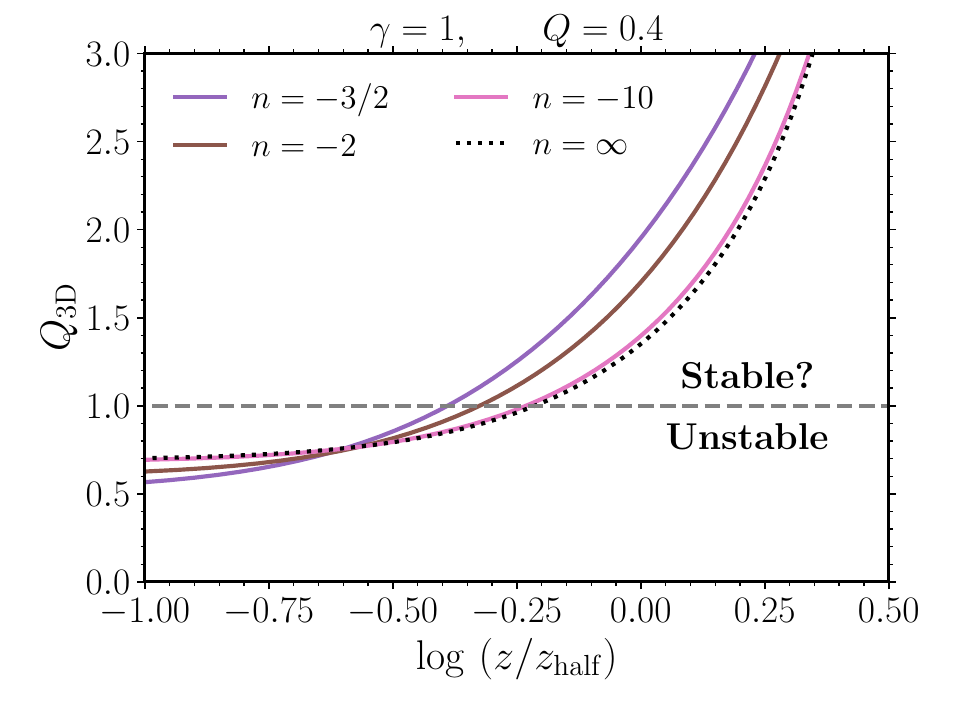}\\
  \caption{Vertical profiles of the local gravitational instability
    parameter $\Qthreed$ for discs with polytropic vertical density
    distributions assuming $Q=0.4$, when $\gamma=5/3$ (upper panel) or
    $\gamma=1$ (lower panel).  In each panel the selected values of
    $n$ are such that the distribution is convectively stable for the
    assumed $\gamma$.  The case $n=\infty$ is the SGI, for which
    $\Qthreed(z)$ is analytic. The horizontal dashed line indicates
    the instability threshold $\Qthreed=1$.}
    \label{fig:poly_q3d}
\end{figure}

{We discuss here the stability of the self-gravitating discs based
  on the 3D instability criterion (\Eq\ref{eq:suff_instab_thick}).}
When $\gamma=5/3$ we explore polytropic distributions with polytropic
exponent in the range $0< \gammap\leq 5/3$ (i.e.\ $n\geq 3/2\,\cup\,
n<-1$). When $\gamma=1$ we explore polytropic distributions with
polytropic exponent in the range $0< \gammap\leq 1$ (i.e.\ $n<-1$).
With these choices, the systems are guaranteed to be convectively
stable.  Fig.\ \ref{fig:poly_q3d} shows vertical $\Qthreed$ profiles
for models with $Q=0.4$ for a selection of values of $n$ for
$\gamma=5/3$ (upper panel) and $\gamma=1$ (lower panel).  The
$\Qthreed$ profiles are qualitatively similar for all values of
$\gamma$ and $n$, with instability close to the midplane and
$\Qthreed$ increasing with $z$. The instability strip $|z|<\zinst$
(see Section \ref{sec:stabunresp}) tends to be slightly wider for
higher $n$, though $\zinst/\zhalf$ spans a small range: $0.35\lesssim
\zinst/\zhalf\lesssim 0.54$ for $Q=0.4$.  {The value $Q=0.4$
  adopted in Fig.\ \ref{fig:poly_q3d} is such to have an unstable
  region around the midplane for all the explored values of $n$. The
  effect of increasing $Q$ is to ``shift upwards'' the curves in
  Fig.\ \ref{fig:poly_q3d}, thus decreasing the extent of the unstable
  regions, which gradually shrink to zero, starting from polytropic
  models with the lowest positive values of $n$.}

\section{Conclusions}
\label{sec:concl}

Building on the 3D instability analysis of \citet{Nip23}, we have
studied the local gravitational instability in two-component 3D
axisymmetric discs.{We have focused on two-component isothermal
  discs (without vertical velocity dispersion gradients), but we have
  complemented our analysis with one-component self-gravitating
  polytropic discs (with vertical velocity dispersion gradients).} The
main results of this paper are the following.

\begin{itemize}

\item {Given that the effect of a second responsive disc is to favour
  the instability, for each component of a two-component disc
  $\Qthreed<1$ is a sufficient condition for instability.} Under the
  assumption that the vertical distributions are isothermal,
  $\Qthreed$ can be computed for each component at given radius $R$ as
  a function of distance from the midplane $z$ using observational
  estimates of the surface densities, velocity dispersions and of the
  epicycle frequency $\kappa$.

\item At given $R$ and $z$, and at fixed surface density, velocity
  dispersion and $\kappa$ (thus at fixed 2D $Q$ parameter), the 3D
  $\Qthreed$ parameter of a disc can be higher, but can also be lower
  in the presence than in the absence of a second component (see lower
  panel of Fig.\ ~\ref{fig:lowq}). When present, the instability
  occurs in a strip enclosing the midplane with half-height $\zinst$,
  which can be computed numerically from observationally inferred
  quantities.

\item {We derived a sufficient condition for local gravitational
  instability of two-component vertically stratified discs that takes
  into account the mutual back-reactions of the two discs.} For
  instance, a disc consisting of a gaseous disc (index 'gas') and a
  stellar disc (index '$\star$'), with $\Qthreedgas>1$ and
  $\Qthreedstar>1$ at a given $R$, is unstable in the midplane against
  radial perturbations with wavenumber $\kR$, if
  $\Fgasplusstar(\kR)>1$, where $\Fgasplusstar$ is a function that can
  be expressed in terms of observationally inferred quantities (see
  Section~\ref{sec:fgasplustar}).

\item We have computed $\Qthreed(z)$, at given $R$, for one-component
  self-gravitating discs with vertical polytropic distributions and
  thus vertical temperature or velocity-dispersion gradients. For a
  range of values of the polytropic index $n$ corresponding to
  convectively stable configurations, we found a behaviour
  qualitatively similar to discs with vertical isothermal
  distributions ($n=\infty$). For unstable models, the height of the
  instability region increases only slightly with $n$, at fixed $Q$.
  
\end{itemize}

{In conclusion, the exploration of this paper provides support to
  the proposal that $\Qthreed<1$ is a robust sufficient condition for
  local gravitational instability of discs, which depends only weakly
  on the presence of a second component and on the vertical
  velocity-dispersion gradient.}  When $\Qthreed>1$, the local
gravitational instability of a two-component disc can be tested by
applying the wavenumber-dependent criterion
(\ref{eq:Fgasplusstar}). In conclusion, whenever the observational
data allow to reconstruct the 3D properties of discs
\citep[see][]{Bac24}, 3D local gravitational instability criteria such
as those analyzed in this work can be successfully employed.

\section*{Acknowledgements}

{We thank the referee Alessandro Romeo for useful comments that
  helped improve the paper.} The research activities described in this
paper have been co-funded by the European Union – NextGenerationEU
within PRIN 2022 project n.20229YBSAN - Globular clusters in
cosmological simulations and in lensed fields: from their birth to the
present epoch.

%

\bibliographystyle{aa}
\bibliography{biblio_vertq3d.bib}

%
\onecolumn
\begin{appendix}

  \section{Gravitational potential of radial perturbations in thick discs}
\label{sec:pot_rad_pert}
  
When studying radial axisymmetric linear perturbations in vertically
stratified discs, \citetalias{Nip23}, following \citet{Gol65a},
assumed a perturbed Poisson equation in the form
\begin{equation}
 -\left(\kR^2+\hz^{-2}\right)\delta\Phi=4\pi G\deltarho,
\label{eq:poisson_hz}
\end{equation}
which approximately accounts for the finite vertical extent of the
disc. In \Eq(\ref{eq:poisson_hz}), $\delta\Phi$ is the perturbed
potential, $\delta\rho$ is the perturbed density, $\kR$ is the
perturbation wavenumber and $\hz$ is the disc thickness. Following
\citetalias{Nip23}, we assume $\hz=\hseventy$ (\Eq\ref{eq:hseventy}),
so the above equation becomes
\begin{equation}
 -\left(\kR^2+\hseventy^{-2}\right)\delta\Phi=4\pi G\deltarho.
 \label{eq:poisson_h70}
 \end{equation}

In this appendix we assess quantitatively the validity of this
approximation of the perturbed Poisson equation. Let us consider an
axisymmetric gravitational potential perturbation, which in
cylindrical coordinates has the form
\begin{equation}
\delta\Phihat(R,z,t)=f(z)\exp[\im(\kR R-\omega t)]\delta\Phihatzero,
\end{equation}
with $f(z)$ such that $f(0)=1$, $\lim_{|z| \to \infty}f(z)=0$ and
$\delta\Phihatzero=\delta\Phihat(0,0,0)$ constant, and a density
perturbation in the form
\begin{equation}
\delta\rhohat(R,z,t)=g(z)\exp[\im(\kR R-\omega t)]\delta\rhohatzero,
\label{eq:deltarhohat}
\end{equation}
with $g(z)$ such that $g(0)=1$, $\lim_{|z| \to \infty}g(z)=0$ and
$\delta\rhohatzero=\delta\rho(0,0,0)$ constant.  From the perturbed
Poisson equation
\begin{equation}
\nabla^2\delta\Phihat=4\pi G\delta\rhohat,
\end{equation}
assuming (as in the analysis of \citetalias{Nip23}) that $1/R$ is negligible compared to $|\kR|$, we get
\begin{equation}
\frac{1}{\Azero}\left(-\kR^2 f+\frac{\dd^2 f }{\dd z^2}\right)\exp[\im(\kR R-\omega t)] \delta\Phihatzero=\frac{4\pi G}{\Azero} g(z) \exp[\im(\kR R-\omega t)]  \delta\rhohatzero,
\end{equation}
which is satisfied for
\begin{equation}
g(z)=\frac{1}{\Azero}\left(-\kR^2 f+\frac{\dd^2 f }{\dd z^2}\right) 
\label{eq:gz}
\end{equation}
and
\begin{equation}
\delta\Phihatzero=\frac{4\pi G}{\Azero}\delta\rhohatzero,
\label{eq:deltaphihatzero}
\end{equation}
where
\begin{equation}
  \Azero\equiv-\kR^2+\left(\frac{\dd^2 f }{\dd z^2}\right)_{z=0}
  \label{eq:azero}
\end{equation}
is a normalization factor that ensures $g(0)=1$. Indicating with
$\langle\cdots \rangle$ a weighted average in $z$, in our perturbed
hydrodynamics equations we can approximate $\delta\Phihat$ with
\begin{equation}
\delta\Phi (R,t)\equiv\langle \delta\Phihat \rangle
=\langle f \rangle\exp[\im(\kR R-\omega t)]\delta\Phihatzero.
\end{equation}
Similarly, we can approximate
$\delta\rhohat(R,z,t)$ with 
\begin{equation}
\delta\rho (R,t)\equiv \langle \delta\rhohat\rangle=\langle g \rangle\exp[\im(\kR R-\omega t)]\delta\rhohatzero,
\end{equation}
where (see \Eq\ref{eq:gz})
\begin{equation}
\langle g \rangle =\frac{1}{\Azero}\left(  -\kR^2\langle f \rangle+\left\langle\frac{\dd^2 f }{\dd z^2}\right\rangle\right).
\label{eq:gangle}
\end{equation}


Let us specialize to the case in which
\begin{equation}
f(z)=e^{-\frac{z^2}{2 H^2}}
\label{eq:fz}
\end{equation}
and adopt as weighted average of a generic function $F(z)$
\begin{equation}
\langle F(z)\rangle =\frac{\int_{-\infty}^\infty F(z)f(z)\dd z}{\int_{-\infty}^\infty f(z)\dd z}.
\end{equation}
For $f$ given by \Eq(\ref{eq:fz}) $\Azero=-(\kR^2+H^{-2})$
(\Eq\ref{eq:azero}), so, combining \Eq~(\ref{eq:gz}) and
\Eq~(\ref{eq:fz}), we get
\begin{equation}
g(z)=\frac{\kR^2+H^{-2}-z^2H^{-4}}{\kR^2+H^{-2}}e^{-\frac{z^2}{2 H^2}}.
\label{eq:gzH}
\end{equation}
Calculating explicitly the weighted averages over $z$, we get
\begin{equation}
\langle f\rangle=\left. \int f^2(z)\dd z \middle/ \int f(z)\dd z=\frac{1}{\sqrt{2}}\right.
\end{equation}
and
\begin{equation}
\langle g \rangle=\left.\int g(z)f(z)\dd z \middle/\int f(z)\dd z\right. =
\frac{1}{\sqrt{2}(\kR^2+H^{-2})}\left(\kR^2+\frac{1}{2 H^2}\right).
\end{equation}
It follows that
\begin{equation}
\delta\Phi=\frac{1}{\sqrt{2}}\exp[\im(\kR R-\omega t)]\delta\Phihatzero
\end{equation}
and
\begin{equation}
\delta\rho=\frac{\kR^2+\frac{1}{2}H^{-2}}{\sqrt{2}(\kR^2+H^{-2})}\exp[\im(\kR R-\omega t)]\delta\rhohatzero.
\end{equation}
Combining the two above equations we get
\begin{equation}
\delta\rho=-\frac{\kR^2+\frac{1}{2}H^{-2}}{\kR^2+H^{-2}}\frac{\delta\rhohatzero}{\delta\Phihatzero}\delta\Phi,
\end{equation}
which, using \Eq~(\ref{eq:deltaphihatzero}), gives 
\begin{equation}
-\left(\kR^2+\frac{1}{2} H^{-2}\right)\delta\Phi=4\pi G\delta\rho.
\end{equation}
The above equation coincides with \Eq~(\ref{eq:poisson_h70}) if we assume
\begin{equation}
2H^2=\hseventy^2.
\label{eq:assumedH}
\end{equation}
We conclude that, approximating the gravitational potential of the
perturbation with its vertical weighted average, the perturbed Poisson
equation in the form (\ref{eq:poisson_h70}) is valid for a density
perturbations with vertical profile
\begin{equation}
  \frac{\delta\rhohat(R,z,t)}{\delta\rhohat(R,0,t)}=g(z)=
  \frac{\kR^2+H^{-2}-z^2H^{-4}}{\kR^2+H^{-2}}e^{-\frac{z^2}{2 H^2}},
\label{eq:gzbis}
\end{equation}
with $H^2=\hseventy^2/2$.

\begin{figure}
  \centerline{\includegraphics[width=0.5\textwidth]{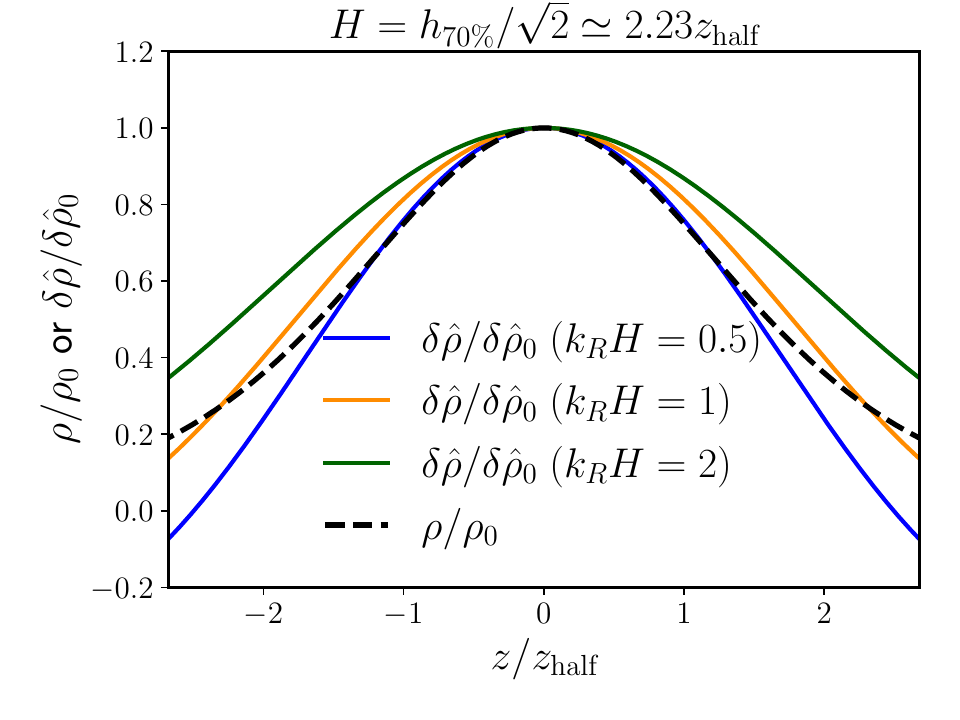}\includegraphics[width=0.5\textwidth]{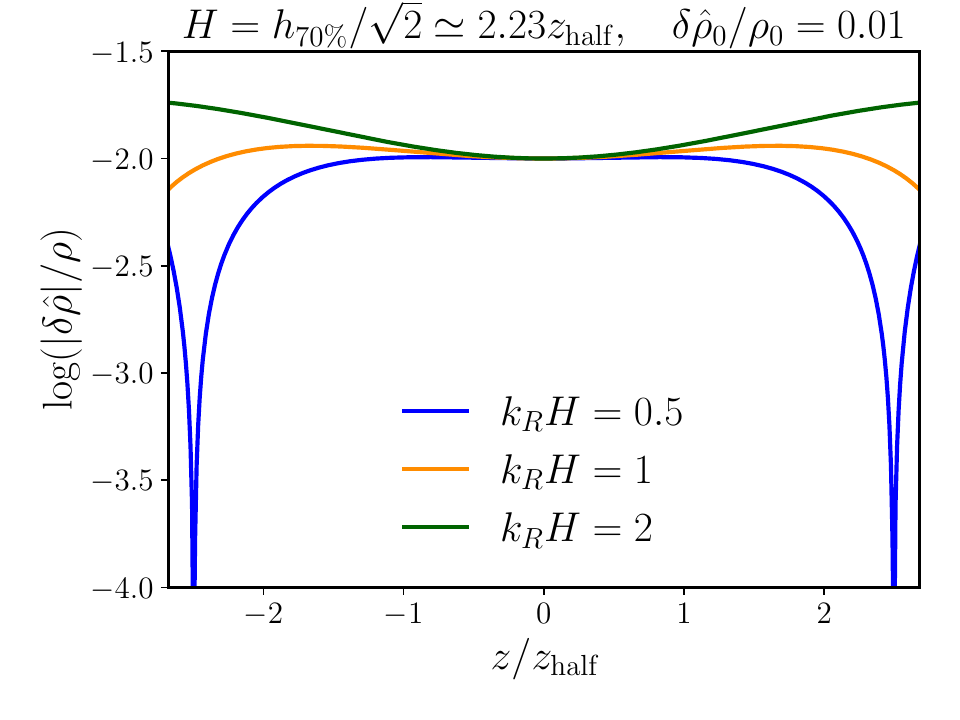}}
  \caption{Left panel. Normalized vertical profiles of the density
    perturbation $\delta\rhohat$ (\Eqs\ref{eq:deltarhohat} and
    \ref{eq:gzH}) with $H=\hseventy/\sqrt{2}$ for different values of
    the radial wavenumber $\kR$ (solid curves), compared with the
    normalized unperturbed vertical density profile
    (\Eq\ref{eq:rho_sgi}; dashed curve).  Right panel: vertical
    profiles of the ratio between the density perturbation and the
    unperturbed density profile for the same values of $\kR$ as in the
    left panel, assuming $\delta\rhohat/\rho=0.01$ at $z=0$.  The
    plotted range in $z$ encloses 90\% of the mass per unit surface of
    the unperturbed distribution.
    \label{fig:deltarhoz}}
\end{figure}

We now want to verify whether assuming a perturbation with this
vertical structure is consistent with the vertical structure of the
unperturbed system.  Let us consider for instance an unperturbed
SGI vertical density profile (\ref{eq:rho_sgi}): in this case
$\hseventy\simeq 1.7346 b$, so the assumption (\ref{eq:assumedH})
becomes
\begin{equation}
H=\frac{1.7346}{\sqrt{2}}b=1.2265 b.
\end{equation}
Assuming $H^2=\hseventy^2/2$, in the left-hand panel of
Fig.~\ref{fig:deltarhoz} we compare, for different values of $\kR H$,
the normalized vertical density profile of the perturbation
$\delta\rhohat(R,z,t)/\delta\rhohat(R,0,t)=g(z)$ (with $g$ given by
\Eq\ref{eq:gzH}) with the normalized unperturbed vertical density
profile $\rho(z)/\rho(0)$. The fact that the shape of the density
profile of the perturbation is similar to that of the unperturbed
density distribution suggests that the considered disturbance can be
consistent with the hypothesis of small perturbations
$|\delta\rho|/\rho\ll 1$. This is illustrated in the right-hand panel
of Fig.~\ref{fig:deltarhoz}, showing the ratio $|\delta\rhohat|/\rho$
as a function of $z$ assuming, for instance, $\delta\rhohat/\rho=0.01$
in the midplane.  We note that in Fig.\ \ref{fig:deltarhoz} the
plotted range in $z$ is such that it contains $90\%$ of the mass per
unit surface of the unperturbed distribution.

\end{appendix}

\end{document}